\documentclass[onecolumn,superscriptaddress,prd,aps,floats,floatfix,nofootinbib,amsmath,amssymb]{revtex4}
\usepackage{graphicx}
\usepackage{color}
\usepackage{amsmath,amssymb,enumerate}
\usepackage{array}

\newcommand{\vev}[1]{\left\langle #1 \right\rangle}
\newcommand{\lessim}{\hspace{0.3em}\raisebox{0.4ex}{$<$}\hspace{-0.75em}\raisebox{-.7ex}{$\sim$}\hspace{0.3em}}

\newcommand{\TeV}{\text{TeV}}
\newcommand{\GeV}{\text{GeV}}

\newcommand{\crit}{{\rm crit}}
\newcommand{\tr}{\mbox{tr}}
\newcommand{\gtwo}{I\kern-.1em I\,}
\newcommand{\be}{\begin{equation}}
\newcommand{\ee}{\end{equation}}
\newcommand{\beq}{\begin{eqnarray}}
\newcommand{\eeq}{\end{eqnarray}}
\newcommand{\bpm}{\begin{pmatrix}}
\newcommand{\epm}{\end{pmatrix}}
\newcommand{\cl}{\, \rm C.L.}

\begin{document}

\title{Vacuum Alignment of the Top-Mode Pseudo-Nambu-Goldstone Boson Higgs Model }
\author{Hidenori S. Fukano}
\thanks{\tt fukano@kmi.nagoya-u.ac.jp}
      \affiliation{ Kobayashi-Maskawa Institute for the Origin of Particles and 
the Universe (KMI) \\ 
 Nagoya University, Nagoya 464-8602, Japan.}
\author{Masafumi Kurachi} \thanks{\tt kurachi@kmi.nagoya-u.ac.jp}
      \affiliation{ Kobayashi-Maskawa Institute for the Origin of Particles and 
the Universe (KMI) \\ 
 Nagoya University, Nagoya 464-8602, Japan.}
\author{Shinya Matsuzaki}\thanks{\tt synya@hken.phys.nagoya-u.ac.jp}
      \affiliation{ Institute for Advanced Research, Nagoya University, Nagoya 464-8602, Japan.}
      \affiliation{ Department of Physics, Nagoya University, Nagoya 464-8602, Japan.}

\begin{abstract}
We study the vacuum alignment 
of the top-mode pseudo-Nambu-Goldstone boson Higgs (TMpNGBH) model,  
which has recently been proposed as a variant of the top quark condensate model 
in light of the 126 GeV Higgs boson discovered at the LHC.  
It is shown that the vacuum of the model, 
determined from the one-loop effective potential with all the explicit breaking effects included, 
realizes the  electroweak symmetry breaking with the appropriate breaking scale. 
Phenomenologies of two characteristic particles in the TMpNGBH model, 
namely the $CP$-odd partner of the Higgs ($A^0_t$) 
and the vectorlike partner of the top quark ($t'$) 
are also studied based on the newly identified vacuum.   
\end{abstract}

\maketitle

%%%%%%%%%%%%%%%%%%%%%%%%%%%%%%%%%%%%%%%%%%
%%%%%%%%%%%%%%%%%%%%%%%%%%%%%%%%%%%%%%%%%%
%%%%%%%%%%%%%%%%%%%%%%%%%%%%%%%%%%%%%%%%%%
%%%%%%%%%%%%%%%%%%%%%%%%%%%%%%%%%%%%%%%%%%
\section{Introduction}
\label{sec-intro}

The discovery of a $126\,\GeV$ Higgs boson at the LHC~\cite{Aad:2012tfa,Chatrchyan:2012ufa} 
implies that the era to reveal the origin of mass of the elementary particles has come. 
Preceding the discovery of the Higgs boson by about two decades 
the top quark has been discovered at the Tevatron~\cite{Abe:1995hr,Abachi:1995iq}.
The top quark is the heaviest particle among the observed particles 
and its mass is $m_t \simeq 173 \,\GeV$~\cite{Agashe:2014kda}, 
which is coincidentally on the order of the Higgs mass and  
the electroweak symmetry breaking (EWSB) scale 
($v_{_{\rm EW}} \simeq 246\,\GeV$). 
Considering such observed coincidence, 
it is worth considering a scenario in which the top quark 
plays a crucial role to explain the dynamical origin for both the EWSB and the Higgs boson. 

The top quark condensate model~\cite{Miransky:1988xi,Miransky:1989ds,Nambu:1989jt,
Marciano:1989xd,Marciano:1989mj,Bardeen:1989ds} is one of such scenarios. 
However, 
the original top quark condensate model is somewhat far from a realistic situation: 
the predicted value of the top quark mass is too large compared with the experimental value. 
In addition, a Higgs boson predicted as a $t \bar{t}$ bound state 
has the mass in a range of $m_t \lessim m_H \lessim 2 m_t$, 
which cannot be identified with the $126\,\GeV$ Higgs boson at the LHC. 

Recently,
a variant class of the top quark condensate model was 
proposed~\cite{Fukano:2013aea,Cheng:2013qwa}. 
In these models 
the realistic top quark mass is obtained by  
the top-seesaw mechanism as in the literature~\cite{ %
Dobrescu:1997nm,Chivukula:1998wd,He:2001fz,Fukano:2012qx,Fukano:2013kia}, 
while a composite Higgs boson emerges as 
a pseudo Nambu--Goldstone boson (pNGB) associated with 
the spontaneous breaking of a global symmetry, 
therefore it is light to be identified as the LHC Higgs boson. 
The model in \cite{Fukano:2013aea} is called the 
Top-Mode pseudo Nambu-Goldstone Boson Higgs (TMpNGBH) model, 
which is outlined as follows.  
It is constructed from 
the top and bottom quarks $q=(t,b)$ and a vectorlike $\chi$ quark, 
a flavor partner of the top quark having 
the same SM charges as those of the right-handed top quark, 
which form a four-fermion interaction:
\beq
{\cal L}_{4f}
=
G_{4f} (\bar{\psi}^i_L \chi_R)(\bar{\chi}_R \psi^i_L)
\,,\label{Lag-4f0}
\eeq
where 
$\psi^i_L \equiv (t_L , b_L, \chi_L)^{T\,i}\,\,( i = 1,2,3)$. 
This four-fermion interaction possesses the global symmetry $G =U(3)_{L} \times U(1)_R$.
When the value of $G_{4f}$ is large enough to form a fermion-bilinear condensate, 
namely $G_{4f}> G_\crit = 8\pi^2/(N_c \Lambda^2)$ 
with $N_c$ being the number of QCD color and $\Lambda$ the cutoff scale of the theory, 
the global symmetry is spontaneously broken down to $H=U(2)_L \times U(1)_V$.
In association with the symmetry breaking, 
the five NGBs emerge as bound states of the $t$ and $\chi$ quarks,  
in addition to a composite heavy scalar boson, corresponding to the $\sigma$ mode of 
the usual Nambu-Jona-Lasinio (NJL) model~\cite{Nambu:1961tp}. 
Three of these five NGBs are eaten by the electroweak gauge bosons 
when the subgroup of $G$ is gauged by the electroweak symmetry 
(and if the condensate is formed in a direction where the electroweak symmetry is broken).   
The other two remain as physical states, 
and they  obtain their masses by additional interaction terms 
which explicitly break 
the global $G=U(3)_{L} \times U(1)_R$ symmetry:
\beq
{\cal L}_h
=
-
\left[\Delta_{\chi \chi} \bar{\chi}_R \chi_L
+ \text{h.c.}
\right]
- G' \left( \bar{\chi}_L \chi_R \right) \left( \bar{\chi}_R \chi_L \right)
\,. \label{Lag-h0}
\eeq 
Then two NGBs become pNGBs, dubbed as top-mode pNGBs (TMpNGBs). 
One of the TMpNGBs, 
which is the $CP$-even scalar ($h^0_t$), 
is identified as the $126\,\GeV$ Higgs boson discovered at the LHC, 
while the other is the $CP$-odd scalar ($A^0_t$), 
which is similar to $CP$-odd Higgs in many models like 
the minimal supersymmetric standard model, 
the two-Higgs doublet model, etc.

Furthermore, 
the model includes another four-fermion interaction term,
\beq
{\cal L}_t = 
G'' \left( \bar{\chi}_L \chi_R \right) \left( \bar{t}_R \chi_L\right) + \text{h.c.} 
\,. 
\label{Lag-t}
\eeq
This, combined with 
Eq.(\ref{Lag-4f0}), generates the top quark mass via the top-seesaw mechanism. 
Note that this term also explicitly breaks the $G$-symmetry, 
but does not contribute to the TMpNGBs' masses ($m_{h^0_t}$ and $m_{A^0_t}$) 
at the leading order.     
However, it was shown that at the next-to-leading order, 
the term in Eq.(\ref{Lag-t}) gives large corrections to the masses 
of $h^0_t$ and $A^0_t$ 
via the top and $\chi$-quark loops~\cite{Fukano:2013aea}.
This, 
namely the fact that even a small explicit breaking term causes 
large correction to physical quantities at the loop level, 
poses a question: 
is the vacuum alignment stable at the loop level ? 
This is the main question we address in this paper.

If there was no explicit breaking term, 
the vacuum associated with the global symmetry breaking 
by the four-fermion interaction in Eq.(\ref{Lag-4f0}) 
is infinitely degenerate. 
The question is, 
which specific point in the degenerate vacua is chosen 
as the true vacuum after all the explicit breaking terms 
(${\cal L}_h$ in Eq.(\ref{Lag-h0}), ${\cal L}_t$ in Eq.(\ref{Lag-t}), 
and electroweak gauge interactions) are turned on.
In \cite{Fukano:2013aea}, 
the vacuum alignment problem was discussed simply 
by looking at the tree level Lagrangian: 
in that case, only the relevant term is ${\cal L}_h$, 
and therefore a proper choice of values of $\Delta_{\chi\chi}$ and $G'$ gives a vacuum 
which breaks the electroweak symmetry appropriately. 
However, at the one-loop level, all the explicit breaking terms will participate 
in determining the effective potential, 
and 
it could potentially destabilize the EWSB vacuum which was fixed at the leading order. 
To see whether the EWSB vacuum is chosen as desired even at the loop level, 
in this paper, 
we derive the effective potential of the TMpNGBs 
at the one-loop level with all the explicit breaking effects included. 
We show that the vacuum alignment is controlled by a single parameter, $\theta_h$, 
which is expressed in term of model parameters, 
and there actually exist parameter choices 
where various phenomenological requirements are satisfied.

This paper is organized as follows. 
In Sec.~\ref{sec-eff-Lagrnagian}, 
we first derive a low-energy effective Lagrangian induced from the TMpNGBH model, 
then the one-loop effective potential is derived. 
In Sec.~\ref{sec-vacuum-alignment}, 
we discuss the vacuum alignment problem based on 
the effective potential which includes contributions from 
the one-loop diagrams of all the SM gauge bosons, fermions and the TMpNGBs. 
Then, we show an example of a set of parameter choice 
which reproduces various physical quantities, 
including the EWSB scale, top quark mass, electroweak precision parameters, and Higgs mass. 
In Sec.~\ref{sec-phenomenologies-A-tprime}, 
we discuss implications for collider phenomenology based on the newly identified true vacuum. 
Sec.~\ref{summary} is devoted to the summary of the paper. 
In appendix \ref{app-BFM}, 
we present the detailed derivation of the one-loop effective Lagrangian 
based on the background field method, 
and 
in appendix \ref{t-prime-coupling}, 
the coupling property and the partial decay widths of $t'$ quark are summarized.

%%%%%%%%%%%%%%%%%%%%%%%%%%%%%%%%%%%%%%%%%%
%%%%%%%%%%%%%%%%%%%%%%%%%%%%%%%%%%%%%%%%%%
%%%%%%%%%%%%%%%%%%%%%%%%%%%%%%%%%%%%%%%%%%
%%%%%%%%%%%%%%%%%%%%%%%%%%%%%%%%%%%%%%%%%%
\section{Effective Lagrangian of the TMpNGBH model}
\label{sec-eff-Lagrnagian}

For the purpose of discussing the vacuum alignment 
of the TMpNGBH model~\cite{Fukano:2013aea}, 
in this section  
we derive an effective Lagrangian described by the TMpNGBs ($h^0_t$ and $A^0_t$), 
the $t'$ quark, the SM gauge bosons and fermions, including terms explicitly breaking 
the global $U(3)_L \times U(1)_R$ symmetry. 

We start from the Lagrangian defined at a cutoff scale $\Lambda$. 
The Lagrangian is constructed from 
the third-generation quarks in the SM, $q=(t,b)$, 
and an $SU(2)_L$-singlet vectorlike quark $(\chi)$ with the hypercharge $+2/3$, 
which are embedded in the $U(3)$-flavor multiplets as  
$\psi^i_{L,R} \equiv (q_{L,R}, \chi_{L,R})^{T\,i}\, ( i = 1,2,3)$, 
as well as the electroweak gauge bosons in the SM:   
\beq
{\cal L}({\Lambda}) 
=
\bar{\psi}_L i \gamma^\mu \partial_\mu \psi_L
+
\bar{q}_R i \gamma^\mu \partial_\mu q_R
+
\bar{\chi}_R i \gamma^\mu \partial_\mu \chi_R
+
{\cal L}_{4f}
+
{\cal L}_h 
+
{\cal L}_{_{\rm EW}} 
+
{\cal L}_t
\,,\label{start-Lag}
\eeq
where 
\beq
{\cal L}_{4f}
&=&
G_{4f} (\bar{\psi}^i_L \chi_R)(\bar{\chi}_R \psi^i_L)
\,,\label{Lag-4f}
\\[1ex]
{\cal L}_h
&=&
-
\left[\Delta_{\chi \chi} \bar{\chi}_R \chi_L
+ \text{h.c.}
\right]
- G' \left( \bar{\chi}_L \chi_R \right) \left( \bar{\chi}_R \chi_L \right)
\,,\label{Lag-h}
\\[1ex]
{\cal L}_{_{\rm EW}}
&=& 
-\frac{1}{4} W^{\hat{a} \mu \nu}W^{\hat{a}}_{\mu\nu}
-\frac{1}{4} B^{\mu \nu}B_{\mu\nu}
+
\bar{\psi}_L \gamma^\mu L_\mu \psi_L
+
\bar{\psi}_R \gamma^\mu R_\mu \psi_R
\,,\label{Lag-EW}\\[1ex]
{\cal L}_t
&=&
G'' \left( \bar{\chi}_L \chi_R \right) \left( \bar{t}_R \chi_L\right) + \text{h.c.} 
\,.\label{Lag-exp-top}
\eeq
The left- and right-gauge fields $L_\mu$ and $R_\mu$ include 
the $SU(2)_L$ and $U(1)_Y$ gauge fields $W_\mu$ and $B_\mu$  
with the gauge couplings $g$ and $g'$ as 
\beq
L_\mu
=
g  W^{\hat{a}}_\mu 
\left(
\begin{array}{cc|c}
&&0\\
\mbox{\raisebox{1.5ex}{\smash{\large$\mspace{5mu}\tau^{\hat{a}}/2$}}}&&0\\ \hline
\mspace{-20mu}0&\mspace{-20mu}0&0
\end{array}
\right)
+
g' B_\mu \bpm 1/6 & 0 & 0 \\ 0 & 1/6 & 0 \\ 0 & 0 & 2/3 \epm 
\quad , \quad
R_\mu
=
 g' B_\mu \bpm 2/3 & 0 & 0 \\ 0 & -1/3 & 0 \\ 0 & 0 & 2/3 \epm 
\,,\label{EW-fermions}
\eeq 
with $\tau^{\hat{a}}\,(\hat{a}=1,2,3)$ being the Pauli matrices 
and $W^{\hat{a}}_{\mu\nu}$ and $B_{\mu \nu}$ 
the field strengths of the electroweak gauge boson fields $W_\mu^{\hat{a}}$ and $B_\mu$. 
The four-fermion interaction term ${\cal L}_{4f}$ in Eq.(\ref{Lag-4f}) 
(with the fermion-kinetic terms)   
is invariant under the transformation of the global symmetry 
$G=U(3)_{\psi_L} \times U(1)_{\chi_R} \times U(2)_{q_R}$, 
while the terms in ${\cal L}_{h}, {\cal L}_{_{\rm EW}}$ and ${\cal L}_t$ 
explicitly break the global $G$-symmetry: 
the $\Delta$ and $G'$ terms in ${\cal L}_h$ break 
the $G$-symmetry down to $U(2)_{q_L} \times U(1)_V$, 
and $U(2)_{q_L} \times U(1)_{\chi_L} \times U(1)_{\chi_R}$, respectively; 
the electroweak gauge interactions in ${\cal L}_{_{\rm EW}}$ only keep 
the $SU(2)_L \times U(1)_Y$ gauge symmetry 
which are embedded in the $G$-symmetry gauged as in Eq.(\ref{EW-fermions});  
the $G''$ term in ${\cal L}_t$ breaks 
the $G$-symmetry down to $U(2)_{q_L} \times U(1)_{\chi_L} \times U(1)_{t_R=\chi_R}$. 
%%
%\textcolor{red}{
The choice of explicit breaking terms here may look rather arbitrary, 
and there is a possibility that other types of explicit breaking terms appear 
depending on the UV physics above the cutoff scale. 
For example, one could add $G_{t\chi} (\bar{\chi}_L t_R)(\bar{t}_R \chi_L)$, 
which is the only additional four-fermion term 
if we assume the UV physics respects the electroweak symmetry, to the Lagrangian. 
However, this term does not give any contribution to the NJL dynamics 
as far as the coupling is small compared to its critical value.
Anyway, in this paper, we restrict the choice of explicit breaking terms 
as minimal as possible for the purpose of clarifying the mechanism 
of producing the Higgs and top quark masses in this scenario.
%}
%%

We shall momentarily turn off all the explicit breaking terms, i.e., 
${\cal L}_h={\cal L}_{_{\rm EW}}={\cal L}_t=0$,    
and derive an effective Lagrangian generated from the fermion-bubble sum diagrams 
at the leading order of the $1/N_c$-expansion. 
For that purpose, we introduce an $U(3)_L$-triplet auxiliary-field,  
$\Phi^i \sim \bar{\chi}_R \psi^i_L$,  
which can be decomposed as
\beq
\Phi = \frac{1}{\sqrt{2}} U \cdot \vec{\phi}
\,,
\label{aux-field}
\eeq
where $\vec{\phi}$ is a three-component real-vector and 
$U$ is a $3 \times 3$ unitary matrix.  
Using Eq.(\ref{aux-field}) 
we rewrite the Lagrangian at the scale $\Lambda$ as follows: 
\beq
{\cal L}(\Lambda) 
=
\bar{\psi}_L i \gamma^\mu \partial_\mu \psi_L
+
\bar{q}_R i \gamma^\mu \partial_\mu q_R
+
\bar{\chi}_R i \gamma^\mu \partial_\mu \chi_R
-
\left[ \bar{\psi}_L \Phi \chi_R + \text{h.c.} \right]
-
\frac{1}{G_{4f}} (\Phi^\dagger \Phi )
\,.
\eeq
Following the procedure in~\cite{Bardeen:1989ds}, 
we integrate out the quantum-fluctuation fields 
for fermions in the momentum shell 
between the cutoff $\Lambda$ and an infrared scale $\Lambda_\chi$.  
By keeping only the ultraviolet-divergent contributions arising from the fermion loops 
at the one-loop level,  
the resultant Lagrangian then takes the form equivalent to the one generated 
via the fermion-bubble sum diagrams 
at the leading order of the $1/N_c$-expansion: 
\beq
{\cal L}_{\rm eff}(\Lambda_\chi < \Lambda)
&=&
\bar{\psi}_L i \gamma^\mu \partial_\mu \psi_L
+
\bar{q}_R i \gamma^\mu \partial_\mu q_R
+
\bar{\chi}_R i \gamma^\mu \partial_\mu \chi_R
+
\partial^\mu \Phi^\dagger \partial_\mu \Phi
-
y \left[ \bar{\psi}_L \Phi \chi_R + \text{h.c.} \right]
-
V_0(\Phi )
\,,\label{start-Lag-1}
\eeq
where 
\beq
V_0(\Phi ) 
&=&
\frac{1}{Z}\left[ \frac{1}{G_{4f}} - \frac{N_c}{8\pi^2} \Lambda^2\right]  (\Phi^\dagger \Phi )
+ \lambda  (\Phi^\dagger \Phi )^2
\,,\label{start-potential}
\eeq
and 
\beq
Z = \frac{1}{y^2} =\frac{1}{\lambda} = \frac{N_c}{16\pi^2} \ln \frac{\Lambda^2}{\Lambda^2_\chi}
\,. \label{wave-reno}
\eeq
When the four-fermion coupling strength $G_{4f}$ satisfies the criticality condition, 
$G_{4f} > G_\crit(= 8\pi^2/(N_c \Lambda^2))$, without loss of generality, 
we may choose the vacuum expectation value (VEV) of the scalar field $\Phi$ as
\beq 
\vev{\Phi} = 
\frac{f}{\sqrt{2}} 
\bpm
0 \\0 \\ 1\\
\epm 
\,. 
\eeq 
$f$ is determined from the stationary condition, $\delta V/\delta \vev{\Phi} =0$, as 
\beq 
 1 - \frac{G_\crit}{G_{4f}} 
 = \frac{8 \pi^2}{N_c} \frac{f^2}{\Lambda^2} 
 \,. \label{stationary-cond}
\eeq 
When the four-fermion coupling strength $G_{4f}$ satisfies the criticality condition, 
$G_{4f} > G_\crit$, 
the scalar field acquires nonzero VEV, $f\neq 0$, 
which triggers the spontaneous symmetry breaking, 
$G=U(3)_L \times U(1)_R \to H = U(2)_L \times U(1)_V$.  
The real vector $\vec{\phi}$ in Eq.(\ref{aux-field}) 
can then be expressed as  
\beq 
\vec{\phi}
=
\sigma_t \cdot \vec{\varphi} 
\quad \text{with} \quad
 \vec{\varphi} 
 =
\bpm
0 \\ 0 \\ 1 \\
\epm\,,
\label{NJL-vac}
\eeq
with $\sigma_t$ being the real scalar field corresponding to 
the $\sigma$ mode as in the usual NJL model. 
The five NGBs emerge 
in association with the spontaneous breaking of the $G$-symmetry, 
which are parametrized in the unitary matrix $U$ in Eq.(\ref{aux-field}) as 
\beq
U &=& 
\exp \left[
\frac{i}{f} \left( 
\sum_{a=4,5,6,7} 
\pi^a_t \lambda^a 
+ \pi^A_t \Sigma_0 \right) 
\right] 
\,, 
\label{eq:defU}
\eeq
where the Gell-Mann matrices $\lambda^a$ are normalized as 
${\rm tr}[\lambda^a \lambda^b]=2 \delta^{ab}$, and $\Sigma_0$ is defined 
as $\Sigma_0 \equiv \text{diag}(0,0,1)$. 

We may integrate out the $\sigma_t$ field since 
its mass generically becomes as large as the cutoff scale $\Lambda$ 
by the radiative corrections from the $\sigma_t$-potential. 
In that case, we may take the scalar field $\Phi$  in Eq.(\ref{aux-field}) 
to be $(f/\sqrt{2}) U \cdot \vec{\varphi}$ 
to approximate the effective Lagrangian Eq.(\ref{start-Lag-1}) as   
\beq 
{\cal L}_{\rm eff}(\Lambda_\chi < \Lambda)
\approx 
\bar{\psi}_L i \gamma^\mu \partial_\mu \psi_L
+
\bar{q}_R i \gamma^\mu \partial_\mu q_R
+
\bar{\chi}_R i \gamma^\mu \partial_\mu \chi_R
+
\frac{f^2}{2}
\tr \left[ \partial_\mu U^\dagger \partial^\mu U \Sigma_0 \right]
-
\frac{yf}{\sqrt{2}}
\left[ \bar{\psi}_L \left( U \Sigma_0 \right) \psi_R + \text{h.c.} \right]
\,,\label{start-NLSM}
\eeq
where 
we have used $\vec{\varphi}^T\cdot  A \cdot\vec{\varphi} = \tr[A\Sigma_0]$ 
for an arbitrary $3 \times 3$ matrix $A$. 

Let us turn on the explicit breaking terms in 
${\cal L}_h,{\cal L}_{_{\rm EW}}$ and ${\cal L}_t$ 
in Eqs.(\ref{Lag-h}), (\ref{Lag-EW}) and (\ref{Lag-exp-top}). 
As noted in \cite{Fukano:2013aea}, 
it turns out that these explicit breaking terms do not affect 
the criticality and stationary conditions in Eq.(\ref{stationary-cond}).   
Using the same auxiliary field as in Eq.(\ref{aux-field}) and 
neglecting the $\sigma_t$ field,  
we thus find that the effective Lagrangian is modified as  
\beq
{\cal L}_{\rm eff}(\Lambda_\chi < \Lambda) 
&=&
-\frac{1}{4} W^{\hat{a} \mu \nu}W^{\hat{a}}_{\mu\nu}
-\frac{1}{4} B^{\mu \nu}B_{\mu\nu}
\nonumber\\
&&
+
\bar{\psi}_L i \gamma^\mu \partial_\mu \psi_L
+
\bar{q}_R i \gamma^\mu \partial_\mu q_R
+
\bar{\chi}_R i \gamma^\mu \partial_\mu \chi_R
+
\bar{\psi}_L \gamma^\mu L_\mu \psi_L
+
\bar{\psi}_R \gamma^\mu R_\mu \psi_R
\nonumber\\
&&
+
{\cal L}_{\rm eff}(U)
\,,\label{start-eff-Lag}
\eeq
with 
\beq
{\cal L}_{\rm eff}(U)
&=&
\frac{f^2}{2}
\tr \left[ D_\mu U^\dagger D^\mu U \Sigma_0 \right]
-
\tilde{m}_{\chi}
\left[
\bar{\psi}_L {\cal M}_f(U) \psi_R + \text{h.c.}
\right]
\nonumber\\
&&
-
c_1f^2
\tr\left[
U^\dagger \Sigma_0
U \Sigma_0
\right]
+
c_2 f^2
\tr\left[
U \Sigma_0
+
\Sigma_0 U^\dagger
\right]
\,,\label{start-NLsM}
\eeq 
where 
\beq
D_\mu U
=
\left(
\partial_\mu 
- i g \hat{W}_\mu 
+ i g' \hat{B}_\mu 
\right)
U
\quad , \quad
\hat{W}_\mu = \sum^3_{\hat{a}=1} W^{\hat{a}}_\mu \frac{\lambda^{\hat{a}}}{2}
\,,\quad
\hat{B}_\mu = B_\mu \frac{\lambda^0}{2}
\,, \quad 
\lambda^0 
=
\left( 
\begin{array}{ccc} 
1 & 0 & 0 \\ 
0 & 1 & 0 \\ 
0 & 0 &  0
\end{array}
\right) 
\,,\label{covariant-derivative-U}
\eeq 
and 
\beq
{\cal M}_f(U)
&=&
U \Sigma_0
+
\frac{G''}{G_{4f}} \Sigma_0 U \Sigma_1
\quad \text{with} \quad
\Sigma_1
= 
\Sigma_0 \cdot \lambda_4 
= 
\bpm
0 & 0 & 0 \\
0 & 0 & 0 \\
1 & 0 & 0 \\
\epm
\,,\label{fermion-mass-matrix} \\
\tilde{m}_{\chi} 
&=& 
\frac{1}{\sqrt{2}} y f 
= \sqrt{\frac{8 \pi^2}{N_c \ln (\Lambda^2/\Lambda^2_\chi)}} f 
\,.   \label{m-chi}
\eeq
In Eq.(\ref{m-chi}) we have used Eq.(\ref{wave-reno}). 
The coefficients $c_{1}$ and $c_{2}$ in Eq.(\ref{start-NLsM}) are given by 
\beq
c_1 
= 
\frac{y^2}{2} \frac{G'}{G^2_{4f}} 
\,,\quad
c_2
=
\frac{y }{\sqrt{2}f} \frac{\Delta_{\chi \chi}}{G_{4f}}
\,.
\label{def-c1-c2}
\eeq 

Note that, at the tree level, the form of the potential term for NGBs, 
corresponding to the second line of Eq.(\ref{start-NLsM}), 
is determined solely by the ${\cal L}_h$, 
and 
the effect of the explicit breaking terms in ${\cal L}_{_{\rm EW}}$ and ${\cal L}_t$ appear 
only at loop level. 
Therefore, to see the effect of all the explicit breaking terms, 
we compute the effective Lagrangian at one-loop level 
by including all the contributions from the NGBs, electroweak gauge bosons, as well as fermions. 
The effective Lagrangian is calculated by keeping only the quadratic divergent terms, 
and the resultant expression becomes as follows 
(for the detail of the calculation, see Appendix~\ref{app-BFM}):
\beq
{\cal L}^{\text{1-loop}}_{\rm eff} (U)
&=& 
\frac{f^2}{2}
\left( 1 - \frac{\Lambda^2_\chi}{4\pi^2 f^2}\right) 
\tr
\left[
\bar{D}_\mu U^\dagger \bar{D}^\mu U 
\Sigma_0
\right]
-
\tilde{m}_\chi 
\left[
\bar{\psi}_L {\cal M}_f(U) \psi_R + \text{h.c.}
\right]
\nonumber\\
&&
-
\left[
c_1 f^2
\left( 1 - \frac{3 \Lambda^2_\chi}{8\pi^2 f^2}\right) 
-
\frac{f^2 \Lambda^2_\chi}{32\pi^2} 
\left(
 \frac{9}{4} g^2 + \frac{3}{4} g'^2 
+
2N_c y^2 \left( \frac{G''}{G_{4f}}\right)^2
\right) 
\right]
\tr\left[
U^\dagger \Sigma_0
U \Sigma_0
\right]
\nonumber\\
&&
+
c_2 f^2
\left( 1 - \frac{5 \Lambda^2_\chi}{32\pi^2 f^2}\right) 
\tr\left[
U \Sigma_0
+
\Sigma_0 U^\dagger
\right]
\,,\label{1loop-eff-Lag-0}
\eeq 
where the ultraviolet divergences have been cutoff by the cutoff scale of the effective Lagrangian 
$\Lambda_\chi$~\footnote{
Note the same-sign contributions from the gauge and fermion loops in Eq.(\ref{1loop-eff-Lag-0}).  
This counterintuitive result is understood by the fact that 
the gauge-loop contribution arises as the form of 
$\tr[U^\dagger \Sigma_0 U \lambda_0]= - \tr[U^\dagger \Sigma_0 U \Sigma_0] + {\rm constant}$, 
while the fermion-loop as $-\tr[U^\dagger \Sigma_0 U \Sigma_0]$, 
up to the common loop factor. 
See also Eq.(\ref{quad-div-EWloop}). 
}.    
The quadratic divergences 
can be absorbed by redefinitions of the bare coupling $f$, $c_1$ and $c_2$: 
\beq
F^2
&=&
f^2 - \frac{\Lambda^2_\chi}{4\pi^2} 
=  
\frac{N_c}{8\pi^2} \tilde{m}^2_\chi \ln \frac{\Lambda^2}{\Lambda^2_\chi} 
- \frac{\Lambda^2_\chi}{4 \pi^2} 
\,,\label{redef-decayconstant} 
\\ 
C_1 F^2 &=&
c_1 f^2
\left( 1 - \frac{3\Lambda^2_\chi}{8\pi^2 f^2}\right) 
-
\frac{f^2 \Lambda^2_\chi}{32\pi^2} 
\left(
 \frac{9}{4} g^2 + \frac{3}{4} g'^2 
+
2N_c y^2 \left( \frac{G''}{G_{4f}}\right)^2
\right) 
\,, \label{redef-c1} \\ 
C_2F^2 
&=& 
c_2 f^2
\left( 1 - \frac{5 \Lambda^2_\chi }{32\pi^2 f^2}\right) 
\,, \label{redef-c2}
\eeq 
where in Eq.(\ref{redef-decayconstant}) we used Eq.(\ref{m-chi}). 
Then the one-loop effective Lagrangian Eq.(\ref{1loop-eff-Lag-0}) is redefined  
at the scale $\Lambda_\chi$ as  
\beq
{\cal L}^{\text{1-loop}}_{\rm eff}(U; f, c_1, c_2, \tilde{m}_\chi; \Lambda_\chi)  
&\equiv & 
{\cal L}_{\rm eff}(U; F,C_1,C_2, \tilde{m}_\chi; \Lambda_\chi) 
\nonumber \\ 
&=& 
\frac{F^2}{2}
\tr
\left[
\bar{D}_\mu U^\dagger \bar{D}^\mu U 
\Sigma_0
\right]
-
\tilde{m}_\chi
\left[
\bar{\psi}_L {\cal M}_f(U) \psi_R + \text{h.c.}
\right]
\nonumber\\
&&
- 
C_1 F^2 
\tr\left[
U^\dagger \Sigma_0
U \Sigma_0
\right]
+
C_2 F^2 
\tr\left[
U \Sigma_0
+
\Sigma_0 U^\dagger
\right]
\,.\label{1loop-eff-Lag-reno}
\eeq
Thus we read off the effective potential including all the explicit breaking effects 
along with the quadratic divergences at the one-loop level as 
\beq
V_{\rm eff} (U) 
=
C_1 F^2 
\tr\left[
U^\dagger \Sigma_0
U \Sigma_0
\right]
-
C_2 F^2 
\tr\left[
U \Sigma_0
+
\Sigma_0 U^\dagger
\right]
\,.\label{eff-potential-TMP}
\eeq
In the next section we will discuss the vacuum alignment based on 
the effective potential Eq.(\ref{eff-potential-TMP}) 
with the parameters $F$, $C_1$ and $C_2$ defined in 
Eqs.(\ref{redef-decayconstant}), (\ref{redef-c1}) and (\ref{redef-c2}), respectively.

%%%%%%%%%%%%%%%%%%%%%%%%%%%%%%%%%%%%%%%%%%
%%%%%%%%%%%%%%%%%%%%%%%%%%%%%%%%%%%%%%%%%%
%%%%%%%%%%%%%%%%%%%%%%%%%%%%%%%%%%%%%%%%%%
%%%%%%%%%%%%%%%%%%%%%%%%%%%%%%%%%%%%%%%%%%
\section{Vacuum Alignment}
\label{sec-vacuum-alignment} 

In this section, 
we address the vacuum alignment of the TMpNGBH model 
based on the effective potential Eq.(\ref{eff-potential-TMP}). 
We first show that the EWSB is realized as the global minimum of the effective potential, 
then we fix the model parameters of the effective Lagrangian Eq.(\ref{1loop-eff-Lag-reno}) 
by inputing physical quantities and imposing phenomenological constraints.   

%%%%%%%%%%%%%%%%%%%%%%%%%%%%%%%%%%%%%%%%%%
%%%%%%%%%%%%%%%%%%%%%%%%%%%%%%%%%%%%%%%%%%
\subsection{Searching for the minimum}
\label{sec-searchnig-for-minimum}

The vacuum energy can be obtained simply by replacing $U$ in Eq.(\ref{eff-potential-TMP}) 
with the vacuum expectation value $\vev{U}$. 
With appropriate chiral $U(3)_{L,R}$ rotations of fermion fields $\psi_{L,R}$ 
and redefinition of the $\Delta_{\chi\chi}$, 
the vacuum expectation value of $U$ is generically parametrized 
by a single angle parameter $\theta$ as 
\beq
\vev{U} 
=
\bpm
\cos \theta & 0 & \sin \theta \\[1ex]
0 & 1 & 0 \\[1ex]
-\sin \theta & 0 &  \cos \theta
\epm
\,. \label{def-vev-U}
\eeq
The physical interpretation of the parametrization in Eq.(\ref{def-vev-U}) 
can be obtained by considering the vacuum expectation value of the scalar field $\Phi$ 
in Eq.(\ref{aux-field}): 
\beq
\vev{\Phi} 
= \frac{f}{\sqrt{2}}\cdot 
\vev{U} 
\vec{\varphi}
=
\frac{f}{\sqrt{2}} \cdot \left[ 
\cos\theta
\bpm 0 \\ 0 \\ 1 \epm
+
\sin\theta
\bpm 1 \\ 0 \\ 0 \epm
\right]
\,. \label{vev-Phi}
\eeq
%{
%\color{red}
The present parameter $\sin \theta$ 
corresponds to the vacuum misalignment parameter 
in the composite Higgs model, 
e.g. $\epsilon \equiv v/f_\pi$ in the minimal composite higgs model~\cite{Agashe:2004rs}.
%}
%
Therefore, $\sin\theta \neq 0$ ($\cos\theta \neq 1$) is required as a model 
which realizes the EWSB 
and 
we will show that such an EWSB vacuum is actually realized 
as a global minimum of the effective potential in the TMpNGBH model below. 

Taking $U = \vev{U}$ in Eq.(\ref{eff-potential-TMP}) and using $\vev{U}$ in Eq.(\ref{def-vev-U}), 
we have 
\beq
V_{\rm eff}(U = \vev{U})
= 
V(\cos \theta)
\equiv
F^2 \left[ 
C_1 \cdot \cos^2\theta
-
 2 C_2 \cdot \cos\theta 
\right] 
\,,\label{TMP-vac-energy}
\eeq
where $C_1$ and $C_2$ are given in Eqs.(\ref{redef-c1}) and (\ref{redef-c2}). 
It is possible to determine the vacuum alignment by minimizing the above potential energy 
with respect to the alignment parameter $\cos\theta$. 
In the present model, 
we find that the potential energy Eq.(\ref{TMP-vac-energy}) 
is minimized at a nonzero $\theta=\theta_h$ with 
\beq
\cos\theta \Bigg|_{\theta = \theta_h} = \frac{C_2}{C_1} 
\quad \quad
\text{ only if \quad $C_1 >0$ and $\left| \dfrac{C_2}{C_1} \right | < 1$}
\,,\label{V-min-TMP}
\eeq
to realize the desired vacuum in which 
the electroweak symmetry is broken.  

Here, to discuss how generally the above conditions are satisfied, 
let us simplify the expressions of $C_1$ and $C_2$ 
in Eqs.(\ref{redef-c1}) and (\ref{redef-c2}) 
by taking only the large-$N_c$ leading terms:
\beq
C_1\Bigg|_{N_c \gg 1} 
&=& 
\frac{y^2}{2} \frac{G'}{G^2_{4 f}} 
 - \frac{y^2}{2} \frac{N_c\Lambda^2_\chi}{8 \pi^2} \left( \frac{G''}{G_{4f}} \right)^2 
\,,\label{C1-largeNc}\\ 
C_2\Bigg|_{N_c \gg 1} 
&=& 
\frac{y}{\sqrt{2} f} \frac{\Delta_{\chi\chi}}{G_{4f}} 
\,\label{C2-largeNc}. 
\eeq
The first condition for realizing EWSB vacuum, $C_1>0$, 
can be satisfied when $G'>0$ and parameters are chosen 
so that the second term in Eq.~(\ref{C1-largeNc}) is not as large as the first term. 
This situation can be realized in a wide parameter space of the model.  
The second condition for EWSB, namely $|C_2/C_1|<1$, 
can also be easily achieved by making $C_2$ smaller than $C_1$ 
by tuning model parameters: 
Consider the two extreme cases.   
i) $C_1$ gets large;  
ii) $C_1$ becomes small. 
The case i) can be realized 
when the $G''$-term correction becomes negligibly small 
compared to the first term $\propto G'$. 
Then the condition $|C_2/C_1|<1$ goes like $\Delta_{\chi\chi}/\tilde{m}_\chi < G'/G_{4f}$, 
which can easily be achieved by tuning $\Delta_{\chi\chi}$ and $G'$.  
On the other hand, 
the case ii) can take place 
when the cancellation between the first and second terms in Eq.(\ref{C1-largeNc}) 
becomes remarkable, 
then, in this case, 
the condition $|C_2/C_1| < 1$ can be satisfied by 
making $\Delta_{\chi\chi}/\tilde{m}_\chi$ much smaller than $G'/G_{4f}$.
We should note that 
Eqs.(\ref{C1-largeNc}) and (\ref{C2-largeNc}) are only approximate, 
shown just for the purpose of giving a rough idea of 
how generally the conditions for realizing EWSB vacuum are satisfied,
and will not be used for the actual calculations of physical quantities 
in the rest of the paper.
%%%

To check the stability of the vacuum in Eq.(\ref{V-min-TMP}), 
we examine 
the coefficient of the second order derivatives  
of the one-loop effective potential Eq.(\ref{eff-potential-TMP})  
with respect to $\pi^a_t$ around $\theta =\theta_h$. 
Taking into account the parameterization of vacuum expectation value in Eq.(\ref{def-vev-U}), 
we reparametrize the field $U$ as 
\beq
U 
\to 
 \tilde{U} \equiv R(\theta_h)\, U 
\quad \text{with} \quad
R(\theta_h)
=
\bpm
\cos \theta_h & 0 & \sin \theta_h \\[1ex]
0 & 1 & 0 \\[1ex]
-\sin \theta_h & 0 &  \cos \theta_h
\epm
\,. \label{redef-U}
\eeq
The newly defined $U$ here takes the same form as in Eq.(\ref{eq:defU}), 
but now VEV of it is ${\rm diag}(1, 1, 1)$.
For the vacuum in Eq.(\ref{V-min-TMP}) to be stable, 
the following condition has to  be satisfied:
\beq
{\rm eigenvalues\ of\ \  }m^2_{ab}
\left(
\equiv
\left. 
\frac{\partial^2V_{\rm eff}(\tilde{U})}{\partial \pi^a_t \partial \pi^b_t}
\right|_{\theta=\theta_h}
\right)
\ \geq \ 0
\ ,\label{minimum-condition}
\eeq
where $m^2_{ab}\,,(a,b=4,5,6,7,A)$ 
corresponds to the $(a,b)$-element of the mass matrix of the NGBs. 
We find the non-vanishing elements of the NGB mass-squared matrix take the following forms:
\beq
\bpm
m^2_{44} & m^2_{4A} \\[1ex] m^2_{A4} & m^2_{AA}
\epm
=
2 C_1
\times 
\bpm
\cos\theta_h & -\sin\theta_h \\[1ex]
\sin\theta_h & \cos\theta_h
\epm
\bpm
0 & 0 \\[1ex] 
0 & 1
\epm
\bpm
\cos\theta_h & \sin\theta_h \\[1ex]
-\sin\theta_h & \cos\theta_h
\epm
\,,\label{44-AA-mass-2}
\eeq
and
\beq
m^2_{55} 
= 
2 C_1  \sin^2\theta_h
\,.\label{55-mass-2}
\eeq 
The condition in Eq.(\ref{minimum-condition}) thus requires $C_1 \geq 0$.

The massive state in Eq.(\ref{44-AA-mass-2}) is identified as the $CP$-odd scalar $A^0_t$ 
($A^0_t\equiv -\pi^4_t \sin\theta_h + \pi^A_t \cos\theta_h$), 
while that in Eq.(\ref{55-mass-2}) is the CP-even scalar ($\pi_5\equiv h^0_t$), 
dubbed as the ``tHiggs"~\cite{Fukano:2013aea}. 
These masses are related by the alignment parameter $\theta_h$:     
\beq
m^2_{A^0_t} 
&=&
2C_1  
\,,\label{CPodd-TMP-mass}
\\[1ex]
m^2_{h^0_t} 
&=&
2C_1   \sin^2\theta_h 
\nonumber\\
&=& 
m^2_{A^0_t} \sin^2\theta_h
\,.\label{CPeven-TMP-mass}
\eeq 
Other three eigenvalues of mass-squared matrix vanish, which corresponds to 
three massless NGBs ($\pi^{6,7}_t\,,\pi^4_t \cos\theta_h + \pi^A_t \sin\theta_h$).  
These are the would-be NGBs to be eaten by the electroweak gauge bosons.  
It should be noted from Eqs.(\ref{CPodd-TMP-mass}) and (\ref{CPeven-TMP-mass}) 
that the quadratic divergent contributions to masses of TMpNGBs have been 
fully absorbed into the renormalization of the decay constant $F$, 
the coefficient $C_1$ and the alignment parameter $\theta_h$ (or the coefficient $C_2$):  
this implies that the Higgs boson mass 
is controlled by some tuning of the model parameters, 
$G_{4f}$, $\Delta_{\chi\chi}$, $G'$ and $G''$ 
in the original Lagrangian Eq.(\ref{start-Lag}), as will be discussed later.  

To summarize, the TMpNGBH model properly realizes the EWSB in the vacuum characterized 
by the alignment parameter $\theta_h$ in Eq.(\ref{V-min-TMP}), and 
the TMpNGBs, tHiggs ($h^0_t$) and $CP$-odd scalar ($A^0_t$), obtain their  masses 
by the explicit breaking effects, which are  
related each other as in Eqs.(\ref{CPodd-TMP-mass}) and (\ref{CPeven-TMP-mass}). 

Using Eqs.(\ref{V-min-TMP}), (\ref{redef-U}), (\ref{CPodd-TMP-mass}) and (\ref{CPeven-TMP-mass}), 
we rewrite the Lagrangian Eq.(\ref{1loop-eff-Lag-reno}) 
as 
\beq
{\cal L}_{\rm eff}(U;\Lambda_\chi)
=
{\cal L}_{\rm eff}(\tilde{U};\theta_h;\Lambda_\chi)
&=&  
\frac{F^2}{2}
\tr
\left[
\bar{D}_\mu \tilde{U}^\dagger \bar{D}^\mu \tilde{U} 
\Sigma_0
\right]
-
\tilde{m}_\chi
\left[
\bar{\psi}_L {\cal M}_f(\tilde{U}) \psi_R + \text{h.c.}
\right]
\nonumber\\
&&
+ 
\frac{F^2 m_{A^0_t}^2}{2} 
\tr\left[ 
- \left(
\tilde{U}^\dagger \Sigma_0 \tilde{U} \Sigma_0 
\right) 
+ 
\cos\theta_h \left(
\tilde{U} \Sigma_0 + \Sigma_0 \tilde{U}^\dagger 
\right) 
\right]  
\,.\label{1loop-eff-Lag-reno:2}
\eeq 
Below we will fix the model parameters in the Lagrangian Eq.(\ref{1loop-eff-Lag-reno:2}) 
by imposing phenomenological constraints.

%%%%%%%%%%%%%%%%%%%%%%%%%%%%%%%%%%%%%%%%%%
%%%%%%%%%%%%%%%%%%%%%%%%%%%%%%%%%%%%%%%%%%
\subsection{Fixing the model parameters} 
\label{constraint}

We fix the model parameters $F, \tilde{m}_\chi,  G''/G_{4f}, m_{A^0_t}$ and $\theta_h$ 
in the effective Lagrangian Eq.(\ref{1loop-eff-Lag-reno:2})
by five phenomenological inputs. 
To this end, we need expressions of physical quantities in terms of model parameters. 
Those can be obtained from the results shown in Refs. \cite{Fukano:2013aea,Fukano:2014zpa}: 
all we need to do is to replace the bare parameters $f, m_\chi, \theta$ 
in the expressions derived in those references 
with those redefined in the effective Lagrangian Eq.(\ref{1loop-eff-Lag-reno:2}), 
namely, $F$, $\tilde{m}_{\chi}$, and $\theta_h$. 

Three of the physical inputs are chosen to be 
the electroweak scale, the LHC Higgs mass, 
which is identified with the tHiggs $(h^0_t)$ mass in the present model, 
and the top quark mass:  
\beq
 v_{_{\rm EW}} \simeq 246 \,\GeV
 \,, \qquad 
 m_{h^0_t} \simeq 126 \,\GeV 
\,, \qquad 
 m_t \simeq 173 \, {\rm GeV} 
\,. \label{inputs1}
\eeq                                    
As the fourth physical input, 
we take the value of the $T$ parameter~\cite{Peskin:1990zt,Peskin:1991sw} to be~\cite{Ciuchini:2013pca}
\beq 
T \simeq 0.08 \,. 
\eeq
We should note that the $S$ parameter is quite insensitive to the model parameters 
and 
vanishingly small as shown in~\cite{Fukano:2013aea}. 
Therefore, it is not appropriate to use for fixing the model parameters.

Having specified four physical inputs, only one model parameter is remaining to be fixed. 
To fix its value, we use the Higgs signal strengths data measured at the LHC. 
The Higgs signal strengths can be written as a function of a single parameter, 
say $\theta_h$, as in \cite{Fukano:2014zpa}. 
Performing the goodness of fit test by using these Higgs signal strengths data 
for several decay and production categories 
reported from the LHC experiments as done in \cite{Fukano:2014zpa}~\footnote{
The tHiggs couplings to the SM fermions other than the top-quark, 
such as the bottom quark and tau lepton,    
can be incorporated into the effective Lagrangian Eq.(\ref{1loop-eff-Lag-reno:2}) 
by introducing four-fermion interactions in the original Lagrangian Eq.(\ref{start-Lag}), 
which are responsible for the fermion masses~\cite{Fukano:2013aea,Fukano:2014zpa}.   
}, 
we find the $95\%\cl$ constraint on $\cos\theta_h$ to be $\cos\theta_h \gtrsim 0.97$. 
%%%
Therefore, in the following section, we study the LHC phenomenology 
in the parameter space allowed by the $95\%\cl$ constraint on the Higgs signal strength, 
$0.97 \leq \cos\theta_h \lessim1$.
%%
%\textcolor{red}
%{ %
We here comment on an upper bound of $t'$-quark mass 
which is related to the constraint on $\theta_h$.
In the present model, 
the $t'$-quark mass is proportional to $1/\sin\theta_h$ 
taking into account $m_{t'} \sim \tilde{m}_\chi$ 
and $f \sim v_{_{\rm EW}}/\sin\theta_h$ in Eq.(\ref{m-chi}).
This fact implies that future Higgs coupling measurements 
can impose an upper bound on the $t'$-quark mass.
%}

%%
The summary of the actual values of five model parameters 
which realize physical inputs explained above are as follows.
Here, as a reference point, we show those for $\cos\theta_h = 0.97$, 
which is the choice to make the model as non-SM-like as possible. 
\beq
 && F \simeq 1.0 \, \TeV
\,, \qquad 
\tilde{m}_\chi \simeq 1.8 \,\TeV
\,, \qquad 
G''/G_{4f} \simeq 0.7 
\,, \nonumber \\ 
&& 
m_{A^0_t} \simeq 518 \,\GeV
\,, \qquad 
\cos\theta_h \simeq 0.97 
\,.  \label{fixed-parameter}
\eeq 
The amount of tuning among model parameters, 
especially achieving $\cos\theta_h = C_2/C_1\simeq 0.97$, 
can be roughly seen from 
the approximate expressions of $C_1$ and $C_2$ in Eqs.~(\ref{C1-largeNc}) and (\ref{C2-largeNc}). 
Typically, a percent level tuning of parameters are required. 
Once these are fixed, all other physical quantities can be predicted. 
For example, the mass of the $t'$ quark, a vectorlike partner of the top quark arising 
in the mass basis of the $t$ and $\chi$-quarks~\cite{Fukano:2013aea}, 
to be 
\beq 
 m_{t'} \simeq 1.85\,\TeV
 \,. \label{t-prime-mass}
\eeq

Now that the values of the model parameters 
in the effective Lagrangian Eq.(\ref{1loop-eff-Lag-reno:2}) have been fixed, 
we shall next discuss how those values can be realized 
in terms of the parameters in the original four-fermion interaction model in Eq.(\ref{start-Lag}). 
As a reference point, we take the value of $\Lambda_\chi$ twice the mass of $t'$ quark: 
\beq
 \Lambda_\chi \simeq 3.7 \,\TeV
 \,. \label{Lambda_chi}
\eeq
Then the cutoff of the four-fermion dynamics $\Lambda$ is determined 
from Eq.(\ref{redef-decayconstant}) to be 
\beq 
 \Lambda \simeq 7 \times 10^2\,\TeV 
 \,, \label{cutoff_NJL}
\eeq
and the original decay constant $f \simeq 1.2\,\TeV$. 
From the stationary condition Eq.(\ref{stationary-cond}), 
we also estimate the ratio $G_{4f}/G_\crit$ to get 
\beq 
\frac{G_{4f}}{G_\crit} -1 \simeq 7 \times 10^{-4} \,, \label{G4f}
\eeq 
where $G_\crit = 8\pi^2/(N_c \Lambda^2) \simeq ( 140 \,\TeV)^{-2}$. 
Finally, we determine the size of the explicit breaking parameters 
$G'$ and $\Delta_{\chi\chi}$, 
which is the source of the masses of the TMpNGBs.  
From Eqs.(\ref{def-c1-c2}), (\ref{redef-c1}), (\ref{redef-c2}), and 
using the experimental values for the $SU(2)_L$ and $U(1)_Y$ gauge couplings 
(renormalized at the $Z$ boson mass scale) 
$g=0.653, g'=0.358$~\cite{Agashe:2014kda}, 
we find 
\beq
\frac{G'}{G_{4f}} 
\simeq
2 \times 10^{-5}
\,, \qquad
\frac{\Delta_{\chi\chi}}{\tilde{m}_\chi} \simeq 2 \times 10^{-6}
\,.\label{Gp-dchichi-output}
\eeq 

With the choice of parameters given above, 
the TMpNGBH model thus achieves the realistic situation, 
where the electroweak symmetry is broken with an appropriate scale, 
it passes the electroweak precision test, and 
the 126 GeV Higgs arises as the pNGB having 
the coupling property consistent with the LHC Higgs. In addition to the 126 GeV Higgs, 
the model has the $CP$-odd scalar $A^0_t$ and the $t'$ quark with the masses 
in Eqs.(\ref{fixed-parameter}) and (\ref{t-prime-mass}), respectively, 
which are characteristic to the present model. 
The LHC phenomenologies of these particles will be discussed in the next section. 

%%%
We close this section by discussing the pNGB nature of the tHiggs 
for this particular parameter choice. 
Having a rather large value of $G''/G_{4f} (\simeq 0.7)$, 
one might suspect that $G=U(3)_L \times U(1)_R$ 
is no longer a good approximate symmetry of the model, 
and hence the tHiggs ($h^0_t$) cannot arise as the pNGBs. 
This is, however, not the case: 
Using the fixed values of the parameters in Eq.~(\ref{fixed-parameter}), 
one can evaluate the size of corrections to the tHiggs mass in Eq.~(\ref{CPeven-TMP-mass}) 
arising from the $G''$-term at the order of ${\cal O}((G''/G^2_{4f}))$ as  
\beq 
m^2_{h^0_t} \Bigg|_{G''} 
&=& 
y^2
\frac{1}{G_{4f}} 
\left[ 
\frac{G'}{G_{4f}} - \frac{N_c G_{4f} \Lambda^2_\chi }{8 \pi^2} 
\left( \frac{G''}{G_{4f}} \right)^2 
\right] 
\sin^2\theta_h
\nonumber \\ 
&\simeq&
y^2
\frac{1}{G_{4f}} 
\left[ 
\frac{G'}{G_{4f}} - \frac{\Lambda^2_\chi}{\Lambda^2}
\left( \frac{G''}{G_{4f}} \right)^2 
\right] 
\sin^2\theta_h
\nonumber\\
&\simeq& 
y^2
\frac{G'}{G^2_{4f}} \left[ 1 - (0.68) \right] \sin^2\theta_h
\,, \label{mh2_Gpp}
\eeq
%
%{\color{red}
where we have used the NJL criticality condition: 
$N_c G_{4f} \Lambda^2/(8\pi^2) \simeq 1$
to obtain the second line.
%}
%
The first term is controlled by $G'$-term, 
as it should be, as $h^0_t$ being a pNGB, 
and 
the amount of the $G''$-correction to the tHiggs mass 
is numerically smaller than the first term. 
%%
%\textcolor{red}{
We can see, from the expression in the second line of the above equation, 
this situation is made possible by the fact that 
the $\Lambda_\chi$ and $\Lambda$ are chosen to satisfy $\Lambda_\chi/\Lambda \ll 1$ 
so that the absolute value of the second term becomes smaller than the first term. 
This is the condition we needed to satisfy $C_1>1$ in Eq.~(\ref{V-min-TMP}) 
to achieve the desired EWSB vacuum. 
Therefore, as far as the model parameters are chosen to achieve EWSB vacuum, 
the $G$-symmetry is a good approximate symmetry 
to control the mass of the tHiggs 
even if the $G''$-term has comparable strength to $G_{4f}$-term, 
and 
the tHiggs can indeed be regarded as pNGBs. 
%}
We should also mention that the value of $G''/G_{4f}$ can be made smaller 
by taking larger $m_{t'}$ than we fixed here, 
in which case the correction by $G''$-term becomes even smaller than estimated here.
\section{Implications for collider physics}
\label{sec-phenomenologies-A-tprime}

In this section, 
we shall discuss LHC phenomenologies of 
the $CP$-odd TMpNGB ($A^0_t$) and the vectorlike partner of the top quark ($t'$). 
Though in the previous section, 
we have fixed all the model parameters as in Eq.(\ref{fixed-parameter}) 
by five physical inputs as a reference point, 
we will relax the parameter choice 
by allowing the alignment parameter $\cos\theta_h$ taking a value 
in the range of $0.97 \leq \cos \theta_h \lessim1$. 
We should note again that this is the range where 
the coupling property of the tHiggs to SM particles is consistent with the LHC data at $95\%\cl$
For $0.97 \le \cos\theta_h \lesssim 1$ 
the masses of $A^0_t$ and $t'$ monotonically increase from 
$(m_{A^0_t}, m_{t'})=(518\,\GeV, 1.85\,\TeV)$ 
to infinity as $\cos\theta_h \to 1$.  
In this section, 
we thus study the LHC phenomenologies of $A^0_t$ and $t'$ 
with their masses from $(m_{A^0_t}, m_{t'}) =(518 \,\GeV, 1.85 \,\TeV)$ 
to certain heavier mass regions which are considered to be relevant to the LHC.

\begin{figure}[t]
\begin{center}
\begin{tabular}{cc}
{
\begin{minipage}[t]{0.4\textwidth}
\includegraphics[scale=0.8]{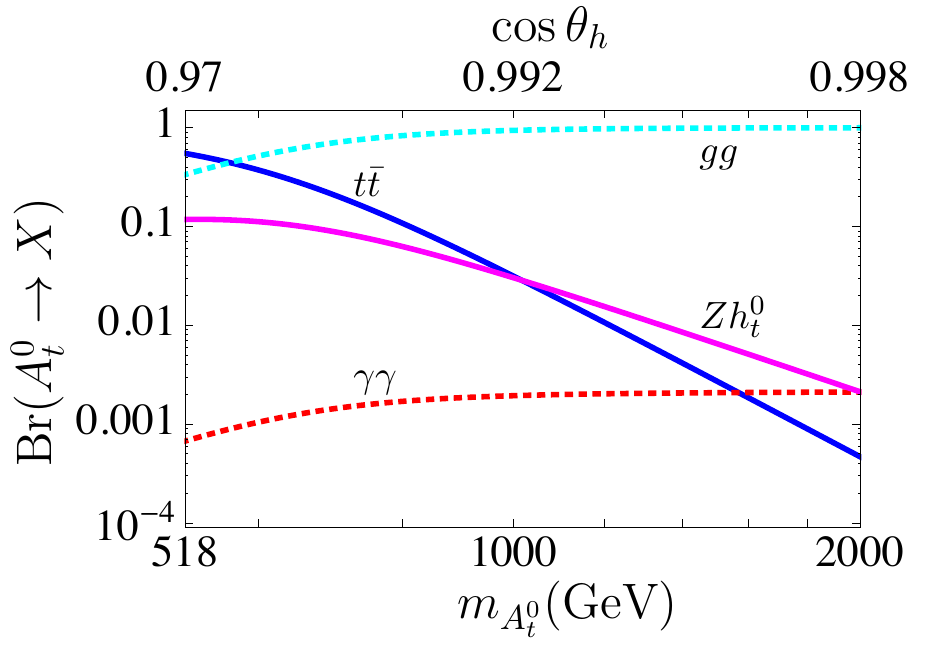} 
%\vspace*{5ex}
\end{minipage}
}
&
{
\hspace{15mm}
\begin{minipage}[t]{0.4\textwidth}
\includegraphics[scale=0.78]{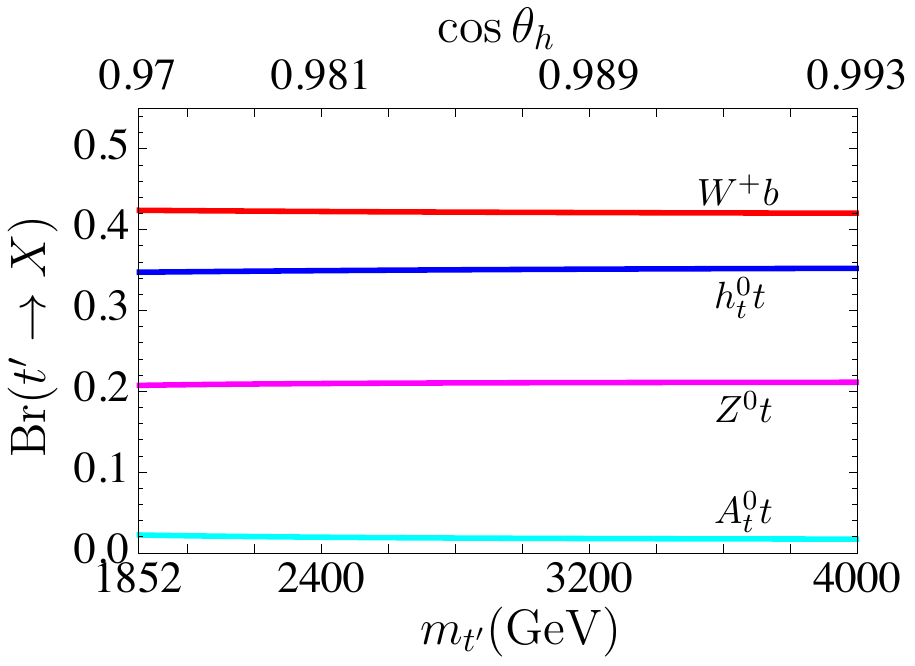} 
%\vspace*{5ex}
\end{minipage}
}
\end{tabular}
\caption[]{
The branching ratios of $A^0_t$ (left panel) and $t'$ (right panel) 
as functions of $m_{A^0_t}$ and $m_{t'}$, respectively. 
Values of $\cos \theta_h$ 
are also shown in the upper horizontal axes. 
\label{Br-A-tprime}}
\end{center}
\end{figure}

The couplings of $A^0_t$ to the SM particles, 
the tHiggs ($h^0_t$) and the $t'$ quark can be read off 
from the Lagrangian Eq.(\ref{1loop-eff-Lag-reno:2}). 
The explicit expressions of the  partial decay widths relevant to the LHC study 
can be found in \cite{Fukano:2014zpa} with the replacement,  
$f \to F$ and $\theta \to \theta_h$. 
In Fig.~\ref{Br-A-tprime}, 
we plot the branching ratio of $A^0_t$ as a function of $m_{A^0_t}$ 
in the range of $518 \,\GeV \leq m_{A^0_t} \leq 2\,\TeV$ 
in the left panel of Fig.~\ref{Br-A-tprime}. 
In this plot, we also indicate the corresponding values of $\cos\theta_h$ 
in the upper horizontal axis. 
This plot is supposed to be the same as the right panel of Fig.2 
in Ref.~\cite{Fukano:2014zpa}, 
though the appearance of the plot looks different, 
especially the branching ratio to $Z h^0_t$ mode. 
The crucial difference between the analysis in Ref.~\cite{Fukano:2014zpa} 
and 
the present one is the presence of the mass relation $m_{h^0_t} = m_{A^0_t} \sin \theta_h$ 
in Eq.(\ref{CPeven-TMP-mass}), 
which is derived consistently at the one-loop level. 

From the plot we see that, in the smaller mass region, 
the $t\bar{t}$ and $gg$ modes 
are the dominant decay channels, 
and therefore the main production process is the gluon-gluon fusion (ggF).   
As was discussed in \cite{Fukano:2014zpa},  
the 8 TeV LHC cross sections 
$pp \to A^0_t \to gg/tt$ for $m_{A^0_t} \geq 1\,\TeV$ 
have not seriously been limited by the currently available data yet 
(\cite{CMS:kxa} for the $gg$-channel 
and 
\cite{TheATLAScollaboration:2013kha,Chatrchyan:2013lca} for the $t\bar{t}$-channel). 
It is therefore to be expected that more data from the upcoming Run-II would probe the $A^0_t$ 
through these channels. 
The detailed study will be  given elsewhere. 
Another interesting channel would be $A^0_t \to Z h^0_t$ as was emphasized 
in the previous analysis \cite{Fukano:2014zpa}.  
However, with the updated branching ratio, 
this channel seems to be rather challenging even 
at the $\sqrt{s}=14\,\TeV$ LHC with 3000 ${\rm fb}^{-1}$ data 
due to the small branching ratio in the smaller mass region.

The $t'$ quark arises as a mixture of the gauge-eigenstate top and $\chi$-quarks 
through the diagonalization of the fermion mass matrix 
in the effective Lagrangian Eq.(\ref{1loop-eff-Lag-reno:2}).   
The explicit expressions of the $t'$ couplings and the partial decay widths 
relevant to the LHC study are listed in Appendix~\ref{t-prime-coupling}. 
In the right panel of Fig.~\ref{Br-A-tprime}, 
we plot the branching ratios of the $t'$ quark as a function of $m_{t'}$. 
In the same way as the plot for the branching ratios of $A^0_t$, 
the corresponding value of $\cos\theta_h$ is also shown in the upper horizontal axis.
From the figure we read off 
\beq
\text{Br}(t' \to W^+ b) \simeq 0.42 
\,, \qquad 
\text{Br}(t' \to Zt) \simeq 0.21 
\,,\qquad 
\text{Br}(t' \to h^0_t t) \simeq 0.35 
\,, \qquad  
\text{Br}(t' \to A^0_t t) \simeq 0.02 
\,. 
\eeq
It is worth comparing these values with the branching ratios of the ``singlet $t'$ quark" 
in a benchmark model of $t'$ quark (e.g.\cite{delAguila:1989rq}), 
$\text{Br}(t' \to W^+ b) \simeq 0.5,  
\text{Br}(t' \to Zt) \simeq 0.25, 
\text{Br}(t' \to h t) \simeq 0.25$, 
for $m_{t'} \simeq 2\,\TeV$~\cite{AguilarSaavedra:2009es,Aguilar-Saavedra:2013qpa}.
It is interesting to note that 
$\text{Br}(t' \to h^0_t t)$ in the present model 
is by about $40\,\%$ larger than that in the benchmark model. 
This is essentially 
due to the large $ht't$ coupling, 
which is the very consequence of the top quark condensate scenario.   

\begin{figure}[t]
\begin{center}
\includegraphics[scale=0.45]{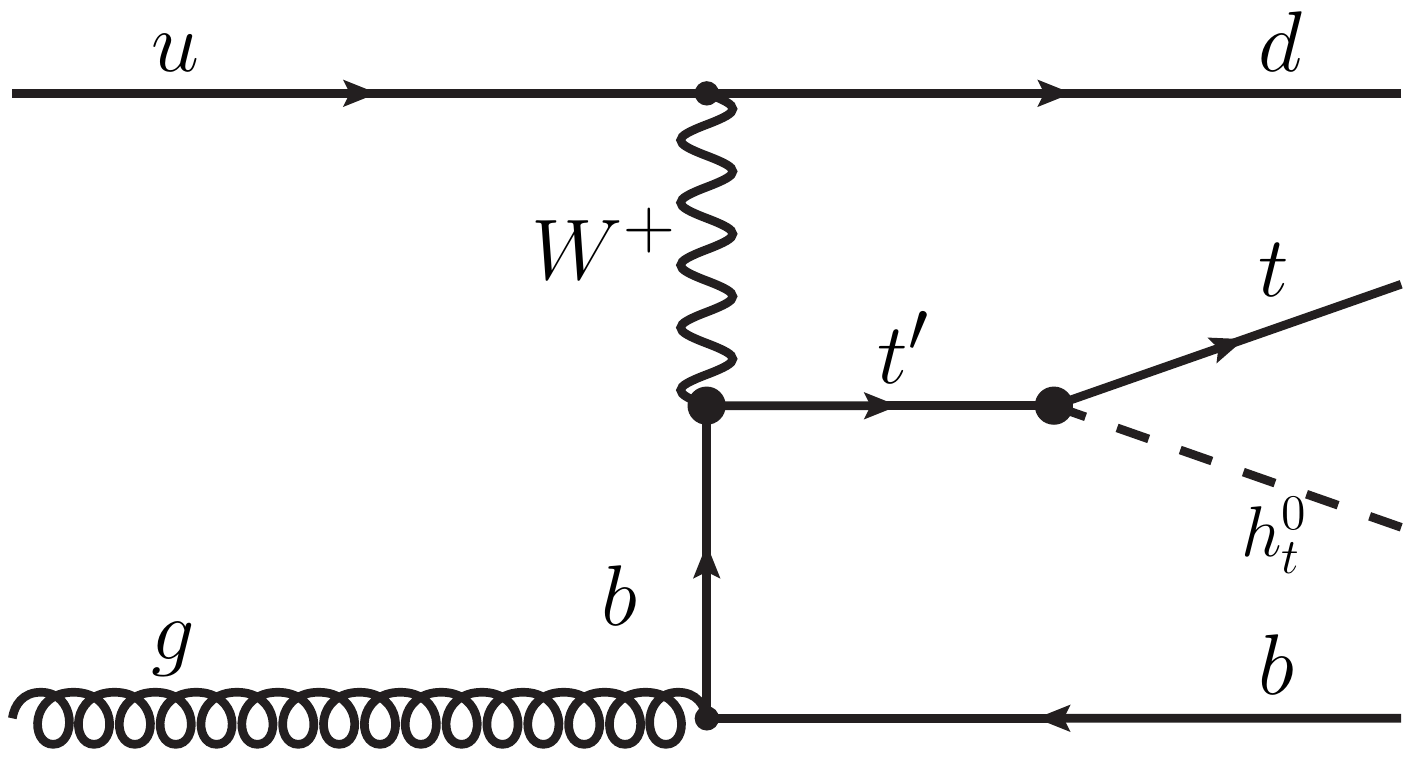}
\caption[]{
An illustration of the single production of $t'$ quark decaying into $h^0_t t$. 
\label{plot-tprime-PH}}
\end{center}
\end{figure}

Since the predicted $t'$ quark in the TMpNGBH model 
is heavy ($m_{t'} \ge 1.8 \,\TeV$), 
it might be challenging to search for the $t'$ quark via the pair production process 
as studied in the usual top-partner search 
by  
the ATLAS \cite{ATLAS:2013ima,TheATLAScollaboration:2013jha,%
ATLAS-CONF-2013-056,TheATLAScollaboration:2013sha,ATLAS-CONF-2014-036} 
and  
the CMS \cite{Chatrchyan:2013uxa,CMS:2014oca,CMS:2014aka,CMS:2014rda} collaborations. 
A more interesting production process of heavy $t'$ quark 
would be the single production, as stressed in %
\cite{AguilarSaavedra:2009es,Aguilar-Saavedra:2013qpa}, 
such as $qg \to t'(\to h^0_t t)\bar{b}j$ as depicted in Fig.\ref{plot-tprime-PH}. 
From Ref.~\cite{Aguilar-Saavedra:2013qpa} 
we can read off the production cross section of the singlet $t'$ quark with $s^t_L \simeq 0.1$ 
(for the definition of this parameter, see Appendix~\ref{t-prime-coupling}),  
$\sigma(t'\bar{t}') \sim  0.1 \,\text{fb}$ 
 at $\sqrt{s}=13\,\TeV$ for $m_{t'} \simeq 2\,\TeV$
in the case of the pair production process, while 
$\sigma(t'\bar{b}j) \sim 4\,\text{fb}$ at $\sqrt{s}=13\,\TeV$ for $m_{t'} \simeq 2\,\TeV$
in the case of the single production. 
By simply quoting these numbers, 
we may roughly estimate the cross sections times the branching ratio 
${\rm Br}(t'\to t h^0_t)\simeq 0.35$ 
in the present model to be
\beq
{\rm pair\ production:\ \  }\sigma(pp \to t'\bar{t}' \to t\bar{t} + 2h^0_t) 
\ &\sim & 
0.01 \,\text{fb} 
\,, \nonumber \\   
{\rm single\ production:\ \ }\sigma(pp \to t'\bar{b}j \to t h^0_t + \bar{b} j) 
&\sim & 
1.4 \,\text{fb} 
\,,  
\eeq
for $\sqrt{s}=13\,\TeV$. 
More details of the $t'$ quark phenomenology at the LHC  
are to be pursued in the future. 

%%%%%%%%%%%%%%%%%%%%%%%%%%%%%%%%%%%%%%%%%%
%%%%%%%%%%%%%%%%%%%%%%%%%%%%%%%%%%%%%%%%%%
%%%%%%%%%%%%%%%%%%%%%%%%%%%%%%%%%%%%%%%%%%
%%%%%%%%%%%%%%%%%%%%%%%%%%%%%%%%%%%%%%%%%%
\section{Summary}
\label{summary}

We addressed the vacuum alignment problem 
of the recently proposed new top quark condensation model \cite{Fukano:2013aea}, 
in which the $126\,\GeV$ Higgs boson emerges 
as a $CP$-even pseudo-Nambu-Goldstone-Boson (TMpNGB)
associated with the global symmetry breaking caused by the supercritical NJL dynamics.
We calculated the one-loop effective Lagrangian for the NGB sector, Eq.(\ref{1loop-eff-Lag-reno}), 
taking into account all the explicit breaking effects, 
including electroweak gauge interactions and four fermion interactions 
responsible for the top-seesaw mechanism. 
The one-loop effective potential Eq.(\ref{eff-potential-TMP}) 
includes all the explicit breaking terms, and therefore 
the correct vacuum is determined by 
the configuration which minimizes the one-loop effective potential. 
It was found that the true vacuum is parameterized 
by $\cos\theta_h$ defined as Eq.(\ref{V-min-TMP}), 
and a non-zero value of $\cos\theta_h$ realizes the EWSB phase 
with the appropriate breaking scale. 

After we clarified relations among physical quantities and model parameters, 
we showed that 
the constraints from the electroweak precision tests 
and the data on the coupling property of the Higgs boson reported by the LHC 
place lower bonds on the masses of the $CP$-odd TMpNGB ($A^0_t$) and $t'$ quark: 
$m_{A^0_t} \gtrsim 520\,\GeV$ and $m_{t'} \gtrsim 1.8\,\TeV$.
We also discussed the collider phenomenologies of $A^0_t$ and $t'$ quark 
on the vacuum aligned at the one-loop level. 
We found that the $A^0_t$ search in the $Z h^0_t$ decay channel would be challenging 
even at the future LHC, 
though the $t\bar{t}$ decay mode is worth investigating. 
As for the $t'$ quark, 
we pointed out that the single production of $t'$ quark, 
decaying into $h^0 t$ at the future LHC could be an interesting discovery channel. 
More detailed study of these collider phenomenologies will be pursued in the future.

%%%%%%%%%%%%%%%%%%%%%%%%%%%%%%%%%%%%%%%%%%
\section*{Acknowledments}
We would like to thank Koichi Yamawaki for useful discussions. 
This work was supported in part by the JSPS Grant-in-Aid for Scientific Research (S) \#22224003.

%%%%%%%%%%%%%%%%%%%%%%%%%%%%%%%%%%%%%%%%%%
%%%%%%%%%%%%%%%%%%%%%%%%%%%%%%%%%%%%%%%%%%
%%%%%%%%%%%%%%%%%%%%%%%%%%%%%%%%%%%%%%%%%%
%%%%%%%%%%%%%%%%%%%%%%%%%%%%%%%%%%%%%%%%%%
\appendix
\section{Background field method}
\label{app-BFM}

In this appendix, 
we derive the one-loop effective Lagrangian Eq.(\ref{start-eff-Lag}) 
based on the background field method. 
The building blocks constructing the effective Lagrangian in Eq.(\ref{start-eff-Lag}) 
are  
\beq 
 U\,, \qquad \hat{W}_\mu\,, \qquad 
\hat{B}_\mu\,, \qquad \psi_{L,R}
\,.
\eeq 
For the purpose of performing the background field method, 
we decompose the above building blocks into the classical fields (denoted by bar fields) 
plus the quantum fluctuating ones (by check fields): 
\beq
U 
&=& 
\bar{U} \cdot \check{U} 
= 
\bar{U} \cdot 
\exp\left[
\frac{i}{f} \left( \sum_{a=4,5,6,7} \check{\pi}^a_t \lambda^a + \check{\pi}^A_t \Sigma_0  \right) 
\right] 
\,,\label{BFM-for-u}\\
\hat{W}_\mu 
&=& 
\bar{W}_\mu + \check{W}_\mu
\,,\label{BFM-for-W}\\
\hat{B}_\mu 
&=& 
\bar{B}_\mu + \check{B}_\mu
\,,\label{BFM-for-B} \\ 
\psi_{L,R} &=& 
\bar{\psi}_{L,R} + 
\check{\psi}_{L,R} 
\,.
\eeq
Then the covariant derivative acting on $U$ given in Eq.(\ref{covariant-derivative-U}) is decomposed as 
\beq
 D_\mu U = (\bar{D}_\mu \bar{U}) \cdot \check{U} + \bar{U} \cdot (\check{D}_\mu \check{U}) 
 \,, 
\eeq
with 
\beq 
 \bar{D}_\mu \bar{U} 
 &=& \partial_\mu \bar{U} - i g \bar{W}_\mu \bar{U} + i g' \bar{B}_\mu \bar{U} 
 \,, \nonumber \\ 
 \check{D}_\mu \check{U} 
 &=& 
 \left( \partial_\mu  - i g \bar{U}^\dagger \check{W}_\mu \bar{U} + i g' \check{B}_\mu \right) \check{U} 
 \,. 
\eeq
And, the gauge field strength for $G_\mu = (\hat{W}_\mu, \hat{B}_\mu)$ is expressed 
in terms of the background and fluctuating fields as 
\beq
 G_{\mu\nu} 
 = 
 \bar{G}_{\mu\nu} 
 + \bar{D}_\mu \check{G}_\nu 
 - \bar{D}_\nu \check{G}_\mu 
 - i g^{(\prime)} [ \check{G}_\mu, \check{G}_\nu ]
 \,, 
\eeq
with 
\beq  
 \bar{G}_{\mu\nu} 
 &=& 
\partial_\mu \bar{G}_\nu - \partial_\nu \bar{G}_\mu - i g^{(\prime)} [\bar{G}_\mu, \bar{G}_\nu]  
\,, \nonumber \\ 
\bar{D}_\mu \check{G}_\nu 
&=& 
\partial_\mu \check{G}_\nu - i g^{(\prime)} [\bar{G}_\mu, \check{G}_\nu] 
\,. 
\eeq

%%%%%%%%%%%%%%%%%%%%%%%%%%%%%%%%%%%%%%%%%%
%%%%%%%%%%%%%%%%%%%%%%%%%%%%%%%%%%%%%%%%%%
\subsection{NG and gauge-boson loops} 

Expanding the Lagrangian Eq.(\ref{start-eff-Lag}) 
in powers of the fluctuating fields up to the quadratic order, 
we find that the NGB and gauge boson sectors take the form 
\beq
{\cal L}_{\rm NGB+EW}
=
{\cal L}^0_{\rm NGB+EW}
+
{\cal L}^{\check{\pi}\check{\pi}}_{\rm NGB+EW}
+
{\cal L}^{\check{G}\check{G}}_{\rm NGB+EW}
+
{\cal L}^{\check{\pi}\check{G}}_{\rm NGB+EW}
+
\cdots
\,,\label{Lag-BFM}
\eeq
where 
\beq
{\cal L}^0_{\rm NGB+EW}
&=&
-\frac{1}{4} \bar{W}^{\hat{a} \mu\nu} \bar{W}^{\hat{a}}_{\mu\nu}
-\frac{1}{4} \bar{B}^{\mu\nu} \bar{B}_{\mu\nu}
\nonumber\\
&&
+ 
\frac{f^2}{2}
\tr
\left[
\bar{D}_\mu \bar{U}^\dagger \bar{D}^\mu \bar{U} 
\Sigma_0
\right]
- c_1f^2
\tr\left[
\bar{U}^\dagger \Sigma_0
\bar{U} \Sigma_0
\right]
+c_2 f^2
\tr\left[
\bar{U} \Sigma_0
+
\Sigma_0 \bar{U}^\dagger
\right]
\,,\label{LBFM-NGBEW-0}\\
{\cal L}^{\check{\pi}\check{G}}_{\rm NGB+EW}
&=&
f \tr \left[
\left( g \check{W}^\mu - g' \check{B}^\mu\right)
\left(
\bar{U} \, \check{\pi}_t \Sigma_0 \, \bar{D}_\mu \bar{U}^\dagger
+
\text{h.c.}
\right)
\right]
+ \frac{f}{2} {\rm tr} \left[
 (D_\mu \check{W}^\mu - D_\mu \check{B}^\mu) 
 \bar{U} \left\{ \check{\pi}_t, \Sigma_0 \right\} \bar{U}^\dagger 
\right] 
\,,\label{LBFM-NGBEW-piG} \\
{\cal L}^{\check{\pi}\check{\pi}}_{\rm NGB+EW}
&=&
\frac{1}{2} \partial^\mu \check{\pi}^a_t \partial_\mu \check{\pi}^a_t
-
\frac{1}{2}
\check{\pi}^a_t \check{\pi}^b_t \sigma^{ab} 
%\nonumber\\
%&&
- 
\frac{1}{2} 
\left( \partial^\mu \check{\pi}^a_t \check{\pi}^b_t - \check{\pi}^a_t \partial^\mu \check{\pi}^b_t \right)
\Gamma^{ab}_\mu
-
\frac{1}{2} 
\left( \partial^\mu \check{\pi}^A _t\check{\pi}^a_t + \check{\pi}^A_t \partial^\mu \check{\pi}^a_t \right)
S^{aA}_\mu
\,,\label{LBFM-NGBEW-pipi}\\
{\cal L}^{\check{G}\check{G}}_{\rm NGB+EW}
&=&
- \tr \left[ 
(\bar{D}_\mu \check{W}_\nu) 
(\bar{D}^\mu \check{W}^\nu) 
- 
% \left( 1 -\frac{1}{\xi} \right)
(\bar{D}^\mu \check{W}_\mu)^2 
\right]
- 
\tr \left[ 
(\bar{D}_\mu \check{B}_\nu) 
(\bar{D}^\mu \check{B}^\nu) 
- 
% \left( 1-\frac{1}{\xi} \right)
(\bar{D}^\mu \check{B}_\mu)^2 
\right]
\nonumber\\
&&
- g \epsilon^{\hat{a}\hat{b}\hat{c}} 
\bar{W}^{\hat{a}}_{\mu\nu}\check{W}^{\hat{b} \mu}\check{W}^{\hat{c} \nu}
+
\frac{f^2}{2}
\tr\left[ 
\left( g \check{W}^\mu - g' \check{B}^\mu\right)
\left( g \check{W}_\mu - g' \check{B}_\mu\right)
\bar{U}
\Sigma_0
\bar{U}^\dagger 
\right]
\,,\label{LBFM-NGBEW-GG}
\eeq
with 
\beq
\Gamma^{ab}_\mu
= 
- \Gamma^{ba}_\mu
&=&
\frac{1}{2}
\tr \left[ 
\bar{U}^\dagger \bar{D}_\mu \bar{U} 
\left(
\lambda^a \Sigma_0 \lambda^b
-
\lambda^b \Sigma_0 \lambda^a
\right)
\right]
\,,\label{def-Gamma-ab}\\
S^{aA}_\mu 
= 
S^{Aa}_\mu
&=&
\frac{1}{2}
\tr \left[ 
\bar{U}^\dagger \bar{D}_\mu \bar{U} 
\left[
\Sigma_0, 
\{
\lambda^a, \lambda^A
\}
\right]
\right]
\,,\label{def-S-ab}\\
\sigma^{ab}
=
\sigma^{ba}
&=& 
\frac{1}{4} \tr\left[
\bar{D}_\mu \bar{U}^\dagger \bar{D}^\mu \bar{U} 
\left(
\left[ 
\lambda^a, \left[ \lambda^b, \Sigma_0\right]
\right]
+
\left[ 
\lambda^b, \left[ \lambda^a, \Sigma_0\right]
\right]
\right)
\right]
\nonumber\\
&&
-
\frac{1}{2} c_1
\tr\left[
\bar{U}^\dagger \Sigma_0
\bar{U} 
\left(
\left[ 
\lambda^a, \left[ \lambda^b, \Sigma_0\right]
\right]
+
\left[ 
\lambda^b, \left[ \lambda^a, \Sigma_0\right]
\right]
\right)
\right]
\nonumber\\
&&
+
\frac{1}{2} c_2
\tr\left[
\{
\lambda^a, \lambda^b 
\}
\left(
\Sigma_0 \bar{U} 
+
\bar{U}^\dagger \Sigma_0 
\right)
\right]
\,.\label{def-sigma-ab}
\eeq 
The quadratic-mixing term between $\check{W}_\mu, \check{B}_\mu$ and $\check{\pi}_t$ 
in the last term of  Eq.(\ref{LBFM-NGBEW-piG}) 
can be eliminated by adding the gauge-fixing term ${\cal L}_{\rm GF}$,  
\beq
{\cal L}_{\rm GF}
=
-\frac{1}{\xi}
\tr \left[ 
\left(
\bar{D}^\mu \check{W}_\mu 
+
\xi \frac{gf}{4} 
\bar{U} 
\{ \check{\pi}_t, \Sigma_0 \}
\bar{U}^\dagger
\right)^2 
\right]
-
\frac{1}{\xi}
\tr \left[ 
\left(
\bar{D}^\mu \check{B}_\mu 
-
\xi \frac{g'f}{4} 
\bar{U} 
\{ \check{\pi}_t, \Sigma_0 \}
\bar{U}^\dagger
\right)^2 
\right]
\,,\label{Lag-GF}
\eeq
with $\xi$ being the gauge-fixing parameter. 

We compute the one-loop corrections arising 
from Eq.(\ref{Lag-BFM}) together with Eq.(\ref{Lag-GF}). 
We work in the Landau gauge $\xi=0$ and focus 
on quadratically divergent contributions.   
In that case, it turns out that 
the ghost term corresponding to the gauge-fixing term Eq.(\ref{Lag-GF}) 
does not contribute to the one-loop order,  
so we can safely drop the ghost contribution. 

From Eqs.(\ref{Lag-BFM}) and (\ref{Lag-GF}) 
we first compute the quadratic-divergent contributions arising from the gauge loops 
and regularize them by the cutoff $\Lambda_\chi$ 
(by extracting the $D=2$ pole in the dimensional regularization) 
to find  
\beq
-
\frac{f^2 \Lambda^2_\chi}{32\pi^2} 
\left( \frac{9}{4} g^2 + \frac{3}{4} g'^2 \right) 
\tr\left[
\bar{U} \Sigma_0 \bar{U}^\dagger \lambda^0 
\right]
\,,\label{quad-div-EWloop}
\eeq 
where we have dropped terms independent of $\bar{U}$. 
\begin{figure}[htbp]
\begin{center}
\begin{tabular}{cc}
{
\begin{minipage}[t]{0.4\textwidth}
\includegraphics[scale=0.4]{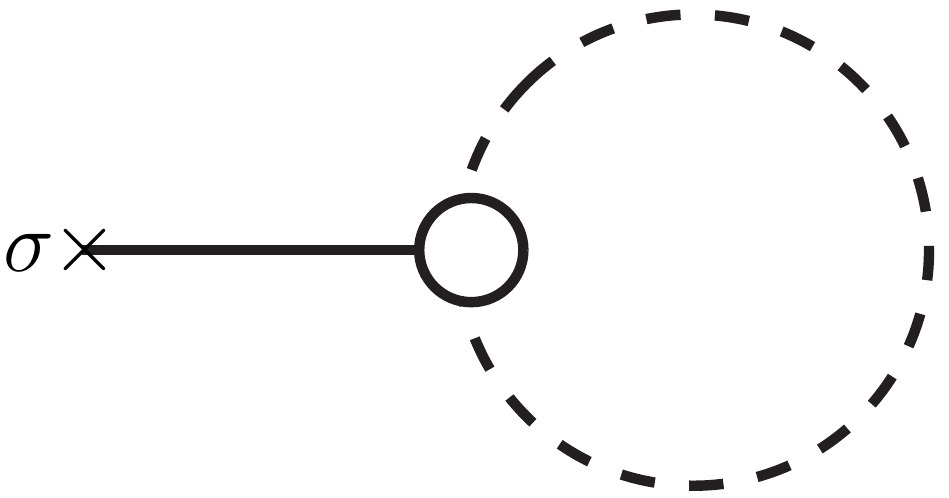} 
%\vspace*{5ex}
\end{minipage}
}
&
{
\begin{minipage}[t]{0.4\textwidth}
\includegraphics[scale=0.4]{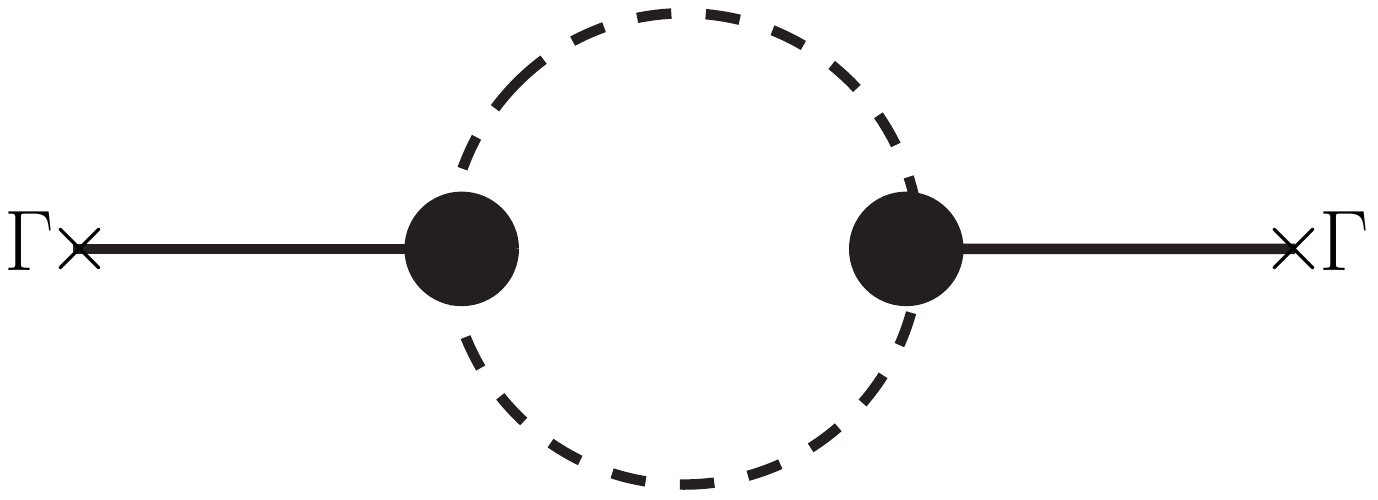} 
%\vspace*{5ex}
\end{minipage}
}
\end{tabular}
\caption[]{
The NGB-loop diagrams giving rise to the quadratic-divergent corrections 
to the effective Lagrangian at the one-loop level in the background field method. 
The dashed line denotes the quantum-fluctuation fields of the NGBs ($\check{\pi}$).
\label{quad-div-NGB}}
\end{center}
\end{figure}%

Evaluating the one-loop diagrams as depicted in Fig.~\ref{quad-div-NGB},  
we next calculate the quadratic-divergent contributions 
from the NGB loops to find  
\beq
\frac{\Lambda^2_\chi}{32\pi^2}
\Biggl[
-4 \tr \left[ \bar{D}_\mu \bar{U}^\dagger \bar{D}^\mu \bar{U} \Sigma_0 \right]
+
12 c_1
\tr\left[
\bar{U}^\dagger \Sigma_0
\bar{U} \Sigma_0
\right]
-
5 c_2 
\tr\left[
\bar{U} \Sigma_0
+
\Sigma_0 \bar{U}^\dagger
\right]
\Biggr]
\,,\label{quad-div-NGBloop}
\eeq
where we have dropped terms independent of $\bar{U}$.  
In reaching Eq.(\ref{quad-div-NGBloop}), 
we have used the following formulae: 
\beq
\sum_{a}
\tr\left[ \lambda^a (A\Sigma_0) \lambda^a (B)\right]
&=&
\frac{1}{2} \tr\left[ BA \Sigma_0 \right]
-
\frac{3}{2}\tr\left[ A \Sigma_0 B \Sigma_0 \right]
+
2 \tr\left[ A \Sigma_0 \right] \tr\left[ B \right]
\,,\\
\sum_{a}
\tr\left[ \lambda^a (\Sigma_0 A) \lambda^a (B)\right]
&=&
\frac{1}{2} \tr\left[ AB \Sigma_0 \right]
-
\frac{3}{2}\tr\left[ A \Sigma_0 B \Sigma_0 \right]
+
2 \tr\left[ A \Sigma_0 \right] \tr\left[ B \right]
\,,\\
\sum_{a}
\tr\left[ \lambda^a (A\Sigma_0)\right]\tr\left[ \lambda^a (B)\right]
&=&
2 \tr\left[ BA \Sigma_0 \right]
-
\frac{3}{2}\tr\left[ A \Sigma_0 B \Sigma_0 \right]
+
\frac{1}{2} \tr\left[ A \Sigma_0 \right] \tr\left[ B \right]
\,,\\
\sum_{a}
\tr\left[ \lambda^a (\Sigma_0 A)\right]\tr\left[ \lambda^a (B)\right]
&=&
2 \tr\left[ AB \Sigma_0 \right]
-
\frac{3}{2}\tr\left[ A \Sigma_0 B \Sigma_0 \right]
+
\frac{1}{2} \tr\left[ A \Sigma_0 \right] \tr\left[ B \right]
\,,\\
\sum_{a}
\tr\left[ \lambda^a \lambda^a (A\Sigma_0 B)\right]
&=&
\frac{9}{4} \tr\left[ BA \Sigma_0 \right]
+
\frac{11}{4}\tr\left[ A \Sigma_0 B \Sigma_0 \right]
\,,
\eeq
and
\beq
\tr\left[ A \Sigma_0 \right] \tr\left[ B \Sigma_0\right]
=
\tr\left[ A \Sigma_0  B \Sigma_0\right]
\,,
\eeq
where $A$ and $B$ are arbitrary $3 \times 3$ matrices.

We thus obtain the effective Lagrangian arising from the NGB and gauge boson loops 
at the one-loop level, 
\beq
\left[
{\cal L}_{\rm NGB+ EW}
+
{\cal L}_{\rm GF}
\right]_{\bar{U}}^{\text{1-loop}}
&=& 
\frac{f^2}{2}
\left( 1 - \frac{\Lambda^2_\chi}{4\pi^2 f^2}\right) 
\tr
\left[
\bar{D}_\mu \bar{U}^\dagger \bar{D}^\mu \bar{U} 
\Sigma_0
\right]
\nonumber\\
&&
-
\left[
c_1 f^2
\left( 1 - \frac{3 \Lambda^2_\chi}{8\pi^2 f^2}\right) 
-
\frac{f^2 \Lambda^2_\chi}{32\pi^2} 
\left( \frac{9}{4} g^2 + \frac{3}{4} g'^2 \right) 
\right]
\tr\left[
\bar{U}^\dagger \Sigma_0
\bar{U} \Sigma_0
\right]
\nonumber\\
&&
+
c_2 f^2
\left( 1 - \frac{5 \Lambda^2_\chi}{32\pi^2 f^2}\right) 
\tr\left[
\bar{U} \Sigma_0
+
\Sigma_0 \bar{U}^\dagger
\right]
\,,\label{Eff-Lag-NGBEW}
\eeq
where we  used $\lambda^0 = 1 - \Sigma_0$. 

%%%%%%%%%%%%%%%%%%%%%%%%%%%%%%%%%%%%%%%%%%
%%%%%%%%%%%%%%%%%%%%%%%%%%%%%%%%%%%%%%%%%%
\subsection{Fermion loops}

Expanding Eq.(\ref{start-eff-Lag}) in powers of 
the fluctuating fields for fermions up to the quadratic order, 
we find the interaction term relevant to the one-loop computation, 
\beq
 {\cal L}_f 
 = 
 {\cal L}_{\rm kin}(f) - \frac{y f}{\sqrt{2}} \left[ 
 \bar{\check{\psi}}_L {\cal M}_f(\bar{U}) \check{\psi}_R 
\right] 
+{\rm h.c.} 
\,,
\eeq
where
\beq
{\cal M}_f(\bar{U})
=
\bar{U} \Sigma_0
+
\frac{G''}{G_{4f}} \Sigma_0 \bar{U} \Sigma_1
\,.
\eeq
From this yukawa interaction we see that the quadratic-divergent contributions from 
fermion loops give rise to  
the effective Lagrangian, 
\beq
\frac{N_c \Lambda^2_\chi}{16\pi^2} y^2 f^2 
\tr\left[
{\cal M}_f(\bar{U}) {\cal M}^\dagger_f(\bar{U})
\right]
%\nonumber\\
=
\frac{N_c \Lambda^2_\chi}{16\pi^2} y^2 f^2 \left( \frac{G''}{G_{4f}}\right)^2
\tr\left[
 \bar{U}^\dagger \Sigma_0 
\bar{U} \Sigma_0
\right]
\,,\label{Eff-Lag-top}
\eeq
where in the second equality 
we have used $\Sigma_1 \Sigma_0=0$, 
$\Sigma^\dagger_1 \Sigma_1 = \Sigma_1\Sigma^\dagger_1 = \Sigma_0$ 
and omitted terms independent of $\bar{U}$.

%%%%%%%%%%%%%%%%%%%%%%%%%%%%%%%%%%%%%%%%%%
%%%%%%%%%%%%%%%%%%%%%%%%%%%%%%%%%%%%%%%%%%
\subsection{Total}

Combining Eqs.(\ref{Eff-Lag-NGBEW}) 
and (\ref{Eff-Lag-top}),
we have 
\beq
{\cal L}^{\text{1-loop}}_{\rm eff} (\bar{U})
&=& 
\frac{f^2}{2}
\left( 1 - \frac{\Lambda^2_\chi}{4\pi^2 f^2}\right) 
\tr
\left[
\bar{D}_\mu \bar{U}^\dagger \bar{D}^\mu \bar{U} 
\Sigma_0
\right]
-
\frac{yf}{\sqrt{2}}
\left[
\bar{\bar{\psi}}_L {\cal M}_f(\bar{U}) \bar{\psi}_R + \text{h.c.}
\right]
\nonumber\\
&&
-
\left[
c_1 f^2
\left( 1 - \frac{3 \Lambda^2_\chi}{8\pi^2 f^2}\right) 
-
\frac{f^2 \Lambda^2_\chi}{32\pi^2} 
\left(
 \frac{9}{4} g^2 + \frac{3}{4} g'^2 
+
2N_c y^2 \left( \frac{G''}{G_{4f}}\right)^2
\right) 
\right]
\tr\left[
\bar{U}^\dagger \Sigma_0
\bar{U} \Sigma_0
\right]
\nonumber\\
&&
+
c_2 f^2
\left( 1 - \frac{5 \Lambda^2_\chi}{32\pi^2 f^2}\right) 
\tr\left[
\bar{U} \Sigma_0
+
\Sigma_0 \bar{U}^\dagger
\right]
\,,\label{1loop-eff-Lag-0-app}
\eeq
where the Yukawa terms for the background fields of fermions 
were added. 
Thus Eq.(\ref{1loop-eff-Lag-0}) has been derived.

%%%%%%%%%%%%%%%%%%%%%%%%%%%%%%%%%%%%%%%%%%%
\section{The $t'$ quark couplings and partial decay widths} 
\label{t-prime-coupling}

In this appendix, 
we shall derive the formulas for the partial decay widths 
of the $t'$ quark relevant to the LHC phenomenology described 
in Sec.~\ref{sec-phenomenologies-A-tprime}.  

Examining the Lagrangian Eq.(\ref{start-eff-Lag}), 
we see that the $t'$ quark in the mass basis, $(t')_m$, arises 
as the mixture of the gauge (current) eigenstates $(t,\chi)^T_g$ through 
the orthogonal rotation which diagonalizes the mass matrix of the seesaw type 
keeping $m_t,m_{t'} \geq 0$~\cite{He:2001fz,Fukano:2013aea},  
\beq
\bpm t_{L} \\[1ex] t'_{L} \epm_m
=
\bpm c^t_{L} & -s^t_{L} \\[1ex] s^t_{L} & c^t_{L}\epm
\bpm t_{L} \\[1ex] \chi_{L} \epm_g
\quad , \quad
\bpm t_{R} \\[1ex] t'_{R} \epm_m
=
\bpm -c^t_{R} & s^t_{R} \\[1ex] s^t_{R} & c^t_{R}\epm
\bpm t_{R} \\[1ex] \chi_{R} \epm_g
\,.\nonumber%\label{rotate-ttprime}
\eeq
The mixing-angle parameters $c^t_{L(R)}$ and $s^t_{L(R)}$ can be expanded 
in powers of $G''/G_{4f} (<1)$ to be 
given up to ${\cal O}((G''/G_{4f})^2)$ as~\cite{Fukano:2013aea}
\beq 
c^t_L 
&=& 
\cos \theta_h \left[
1+ \left( \frac{G''}{G_{4f}} \right)^2 \cos^2\theta_h \sin^2\theta_h
\right]
\quad, \quad
s^t_L 
= 
\sin \theta_h \left[ 
1-\left( \frac{G''}{G_{4f}} \right)^2 \cos^4\theta_h 
\right]
\,,\nonumber%\label{def-Lt}
\\
c^t_R
&=& 
1 - \frac{1}{2}\left( \frac{G''}{G_{4f}} \right)^2 \cos^4\theta_h
\quad, \quad
s^t_R
=
\frac{G''}{G_{4f}}\cos^2\theta_h 
\,. 
\nonumber%\label{def-Rt}
\eeq 
Thus the relevant $t'$ quark interaction-terms are read off from 
Eq.(\ref{start-eff-Lag}) as 
\beq 
{\cal L}_{t'}
&=&
\frac{g}{\sqrt{2}} (s^t_L) 
\left( W^+_\mu \bar{t}'_L \gamma^\mu b_L + \text{h.c.} \right)
+
\frac{g}{2\cos\theta_W} (c^t_L s^t_L) Z_\mu \left( \bar{t}_L \gamma^\mu t'_L + \text{h.c.}\right)
\nonumber\\
&&
-
\frac{y}{\sqrt{2}}
\left[
C_{hL} h^0_t \bar{t}'_L t_R 
+
C_{hR} h^0_t \bar{t}_L t'_R 
+
i C_{AL} A^0_t \bar{t}'_L t_R 
+
iC_{AR} A^0_t \bar{t}_L t'_R 
+
\text{h.c.}
\right]
\,,\label{Lag-for-tprime}
\eeq
where $\theta_W$ is the Weinberg angle 
and the coefficients $C$s are given as~\cite{Fukano:2013aea}
\beq
C_{hL} 
&=&
s^t_R \left( s^t_L \cos\theta_h - c^t_L \sin\theta_h \right)
+
\left( \frac{G''}{G_{4f}} \right) c^t_L c^t_R \sin\theta_h
\,,\nonumber\\
C_{hR} 
&=&
c^t_R \left( c^t_L \cos\theta_h + s^t_L \sin\theta_h \right)
+
\left( \frac{G''}{G_{4f}} \right) s^t_L s^t_R \sin\theta_h
\,,\nonumber\\
C_{AL}
&=&
c^t_L s^t_R
-
\left( \frac{G''}{G_{4f}} \right) c^t_L c^t_R 
\,,\nonumber\\
C_{AR}
&=&
- s^t_L c^t_R
-
\left( \frac{G''}{G_{4f}} \right) s^t_L s^t_R
\,. \nonumber
\eeq

From Eq.(\ref{Lag-for-tprime}), 
we thus evaluate the $t'$ quark decay amplitudes to obtain the formulae for 
the relevant partial decay widths, 
\beq
\Gamma(t' \to W^+ b)
&=&
\frac{g^2}{64\pi} (s^t_L)^2 \frac{m^3_{t'}}{M^2_W} 
\left( 1 - \frac{M^2_W}{m^2_{t'}}\right)^2 
\left( 1 + 2 \frac{M^2_W}{m^2_{t'}}\right) 
\,,%\label{width-tprime-Wb}
\nonumber\\
\Gamma(t' \to Z t)
&=&
\frac{g^2}{64\pi c^2_W} (c^t_L s^t_L)^2 \frac{m^3_{t'}}{2 M^2_Z}
\beta\left( \frac{m^2_t}{m^2_{t'}}, \frac{M^2_Z}{m^2_{t'}}\right) 
\left( 
1 - \frac{2m^2_t - M^2_Z }{m^2_{t'}} + \frac{m^4_t - 2M^4_Z + m^2_t M^2_Z}{m^4_{t'}}
\right) 
\,,%\label{width-tprime-Zt}
\nonumber\\
\Gamma(t' \to h^0_t t)
&=&
\frac{y^2}{32\pi} m_{t'}
\beta\left( \frac{m^2_t}{m^2_{t'}}, \frac{m^2_{h^0_t}}{m^2_{t'}}\right) 
\left[
(C^2_{h L} + C^2_{h R}) \left( 1 + \frac{m^2_t + m^2_{h^0_t}}{m^2_{t'}}\right)
+
4 C_{h L}C_{h R} \frac{m_t}{m_{t'}}
\right] 
\,,%\label{width-tprimes-ht}
\nonumber\\
\Gamma(t' \to A^0_t t)
&=&
\frac{y^2}{32\pi} m_{t'}
\beta\left( \frac{m^2_t}{m^2_{t'}}, \frac{m^2_{A^0_t}}{m^2_{t'}}\right) 
\left[
(C^2_{A L} + C^2_{A R}) \left( 1 + \frac{m^2_t + m^2_{A^0_t}}{m^2_{t'}}\right)
-
4 C_{A L}C_{A R} \frac{m_t}{m_{t'}}
\right] 
\,,\nonumber%\label{width-tprimes-At}
\eeq 
where $\beta(x,y) = \sqrt{(1-x-y)^2 - 4xy}$.

%%%%%%%%%%
\bibliography{ref-alignment,ref-BFM,ref-NLsM,ref-chiralsymmetry,ref-compositeHiggs,ref-SMHiggs,ref-EWPT,ref-2HDM,ref-TC,ref-TMSM,ref-PDG-PDF,ref-Higgs-ATLAS,ref-Higgs-CMS,ref-LHC-woSMHiggs,ref-VLQ,ref-top-quark}
\end{document}